\documentclass[
]{revtex4}
\makeatletter
\def\set@curr@file#1{%
  \begingroup
    \escapechar\m@ne
    \xdef\@curr@file{\expandafter\string\csname #1\endcsname}%
  \endgroup
}
\def\quote@name#1{"\quote@@name#1\@gobble""}
\def\quote@@name#1"{#1\quote@@name}
\def\unquote@name#1{\quote@@name#1\@gobble"}
\makeatother

\usepackage{graphicx}
\usepackage{amssymb}
\usepackage{amsmath}
\usepackage{dcolumn}
\usepackage{bm}
\usepackage{epsfig}
\usepackage{color}
\usepackage{datetime}
\usepackage[T1]{fontenc}
\usepackage[utf8]{inputenc}

\newcommand{\bal}{\begin{align}}
\newcommand{\eal}{\end{align}}
\newcommand{\beq}{\begin{eqnarray}}
\newcommand{\eeq}{\end{eqnarray}}
\newcommand{\nneeq}{\nonumber \end{eqnarray}}

\newcommand{\nn}{\nonumber \\}
\newcommand{\es}{& = &}
\newcommand{\rs}{\, = \,}
\newcommand{\ps}{& + &}
\newcommand{\ms}{& - &}
\newcommand{\ts}{& \times &}

\newcommand{\nt}{\nn \ts}
\newcommand{\np}{\nn \ps}
\newcommand{\nm}{\nn \ms}

\newcommand{\tl}{ & \to & }

\newcommand{\cM}{ {\cal M} }
\newcommand{\cH}{ {\cal H} }

\newcommand{\cV}{ {\cal V} }

\newcommand{\cU}{ {\cal U} }
\newcommand{\cL}{ {\cal L} }

\newcommand{\2}{ \, { 1 \over 2} \, }
\newcommand{\4}{ \, { 1 \over 4} \, }

\begin{document}
\title{ Computation of effective front form Hamiltonians 
         for massive Abelian gauge theory }
\author{ Stanis{\l}aw D. G{\l}azek }
\email{stglazek@fuw.edu.pl}
\affiliation{ Institute of Theoretical Physics\\ 
Faculty of Physics, University of Warsaw, Pasteura 5, 02-093
Warsaw, Poland }
\date{ January 17, 2020 }

\begin{abstract}
Renormalization group procedure for effective particles (RGPEP) 
is applied in terms of a second-order perturbative computation 
to an Abelian gauge theory, as an example of application worth 
studying on the way toward derivation of a dynamical connection 
between the spectroscopy of bound states and their parton-model 
picture in the front form of Hamiltonian dynamics. In addition to 
the ultraviolet transverse divergences that are handled using the 
RGPEP in previously known ways, the small-x divergences are 
handled by introducing a mass parameter and a third polarization 
state for gauge bosons using a mechanism analogous to spontaneous 
breaking of global gauge symmetry, in a special limit that simplifies 
the theory to Soper's front form of massive QED. The resulting 
orders of magnitude of scales involved in the dynamics of effective 
constituents or partons in the simplified theory are identified for the 
fermion and boson mass counter terms, effective masses and 
self-interactions, as well as for the Coulomb-like effective interactions 
in bound states of fermions. Computations in orders higher than 
second are mentioned but not described in this article. 
\end{abstract} 
\maketitle

\section{ Introduction }
\label{intro}

Particle theory singularities that are associated with wee partons 
of the parton model of hadrons~\cite{FeynmanPM} or with 
field quanta that carry small kinematic momenta in the front 
form (FF) of Hamiltonian dynamics~\cite{DiracFF}, require 
a renormalization group procedure that is capable of 
simultaneous handling of the ultraviolet and infrared divergences 
in combination with the bound-state problem, which is a 
complex issue~\cite{Wilsonetal}. One way of approaching 
the issue has been proposed recently~\cite{AbelianAPPB} in 
the context of Abelian gauge theory. The idea is to use
a mechanism analogous to spontaneous breaking of global
gauge symmetry~\cite{Higgs1,EnglertBrout} for introducing 
a mass for gauge bosons and to thus regulate the theory in 
the region in question. This article pursues that idea in terms 
of a study of the kind and magnitude of Hamiltonian interaction 
terms it leads to in the effective theories. Our work is carried 
out in a special limit that simplifies the Abelian theory to Soper's 
FF version of massive QED~\cite{Soper}. The theory does not 
include confinement but it does provide examples of effective 
interactions that bind fermions. 

It should be noted that gauge theories with 
Lagrangian densities similar to Soper’s were 
introduced for analysis in the instant form 
(IF)[2] of dynamics a long time 
ago~\cite{Stueckelberg,Matthews,Coester,Salam,Kamefuchi}. 
In the FF of dynamics, Soper’s work was 
followed by Yan’s~\cite{Yan3,Yan4}. For a review of 
more recent works that use massive vector 
bosons as ultraviolet or infrared regulators 
in FF approaches, see~\cite{Hiller} and references 
therein. Soper found that the replacement of 
photons in FF of QED by massive vector bosons 
is quite simple if one considers in addition 
to the vector field $A^\mu$ in gauge $A^+=0$ 
a scalar field $B$ in the manner of 
Stueckelberg~\cite{Stueckelberg}. The Stueckelberg 
formalism was also used in FF calculations of transition 
matrix elements in the Feynman gauge~\cite{LFStueckelberg}. 
Perhaps similar attempts could be undertaken 
also in the non-Abelian theories~\cite{KunimasaGoto}. 
In view of that extensive record, it should be stated 
up front in what way the present study differs 
from the previous ones. We start from a different 
Lagrangian than massive QED and in the manner 
analogous to spontaneous breaking of global gauge 
invariance arrive at Soper’s theory as a helpful 
simplification in a special limit. Subsequently, 
instead of aiming at reproducing or predicting 
observables directly in terms of the degrees of 
freedom that appear in canonical FF Hamiltonian 
in a diverging way, our goal is to compute 
the equivalent effective FF Hamiltonian operators 
that are written in terms of apparently more 
adequate degrees of freedom [18,19]. Computation 
of such Hamiltonians is hoped to eventually lead 
to a sequence of successive approximations for 
relativistic description of strongly bound states 
because they do not diverge as the canonical FF 
Hamiltonians do, see Sec. IV for details. Soper’s 
theory serves as a preliminary illustration of 
the magnitude of terms one has to deal with. 
Little is known at this point regarding extension 
of our approach to non-Abelian theories. However, 
since the mechanism of spontaneous breaking of 
global gauge symmetry serves the purpose of 
regularization and when one lifts the regularization 
the symmetry may be restored, the author hopes that 
the current exercise with Soper’s theory will turn 
out helpful also in studying non-Abelian theories.

To compute effective FF Hamiltonians for the massive Abelian 
gauge theory, we use the renormalization group procedure for 
effective particles (RGPEP), here only applied up to the second 
order in a series expansion in powers of the coupling 
constant~\cite{RGPEP}. The RGPEP stems from the similarity 
renormalization group (SRG) procedure~\cite{GlazekWilson} 
and draws on the double-commutator differential flow equation 
for Hamiltonian matrices~\cite{Wegner}. The method preserves 
boost-invariance of the FF of Hamiltonian dynamics and its
computations are carried out in terms of the quantum fields on 
one light-front hyperplane in space-time. We calculate the mass 
counter terms, effective fermion and boson mass corrections, 
relativistic fermion-anti-fermion interaction terms that correspond 
to the well-known Yukawa or Coulomb potentials and additional 
terms that do not have classical counterparts. 

Section~\ref{classicaltheory} introduces the classical gauge 
theory we consider. The canonical FF version of the theory 
and its quantization are described in Sec.~\ref{canonicalFFhamiltonian}.
The RGPEP is applied in Sec.~\ref{ARGPEP}, where we compute 
the effective fermion and boson self-interactions and relativistic potentials 
in fermion-anti-fermion systems. Section~\ref{spectroscopyandPMpictures} 
discusses the connection between spectroscopy and the parton-model 
picture of bound states in the context of the RGPEP. Detailed plots 
of mass corrections and relativistic potentials are given in Sec.~\ref{plots}. 
The paper is concluded by Sec. \ref{conclusion}. Appendixes
provide details of our notation and the canonical Hamiltonian of 
Soper's theory.
 
\section{ Classical theory }
\label{classicaltheory}

The FF Hamiltonian for the theory we consider 
was recently derived~\cite{AbelianAPPB} from 
the familiar local Lagrangian density~\cite{Higgs1,EnglertBrout,Kibble} 
\beq
\label{cL}
\cL \es \cL_\psi + \cL_A + \cL_{A\phi} - \cV_{\phi} \ , 
\eeq 
where 
\beq
\label{Lpsi}
\cL_\psi 
\es \bar \psi \left[ \left( i \partial_\mu - g A_\mu \right)
                            \gamma^\mu - m \right] \psi \ , \\
\label{LA}
\cL_A 
\es - {1 \over 4} F_{\mu \nu} F^{\mu \nu} \ , \\
\label{Aphi}
\cL_{A \phi} 
\es \left[ (i\partial^\mu - g' A^\mu ) \phi \right]^\dagger 
(i\partial_\mu - g' A_\mu ) \phi  \ , \\
\label{Lphi}
\cV_\phi
\es
-\mu^2 \ \phi^\dagger \phi + {\lambda^2 \over 2} \ 
( \phi^\dagger \phi )^2 \ .
\eeq
Quanta of field $\psi$ will correspond to fermions and quanta 
of field $A$ to transversely polarized gauge bosons. Quanta 
of the phase of scalar field $\phi$ will supply effects 
associated with the longitudinal polarization of massive 
gauge bosons. This section briefly recapitulates derivation 
of the corresponding FF Hamiltonian and presents it in a 
special limit in which it matches the Hamiltonian designed 
by Soper for the FF of massive QED, a long time ago~\cite{Soper}. 
Further literature on the use of FF quantum dynamics can be 
traced through reviews~\cite{Hiller,FFreview1,FFreview2,FFreview3,
FFreview4,FFreview5,FFreview6}.

\subsection{ Gauge symmetry }
\label{gaugesymmetry}

Field $\phi$ in the Lagrangian density of Eq.~(\ref{Aphi})
can be written using its modulus $|\phi| = 
\varphi/ \sqrt{2}$ and phase $g' \theta$~\cite{Kibble},
\beq
\phi  \es  \varphi \ e^{i g' \theta} /\sqrt{2} \ .
\eeq
The density $\cL_{A\phi}$ is a function of 
$\varphi$, $\partial^\mu \varphi$ and $\partial^\mu \theta$\,
\beq
\label{Aphi2}
\cL_{A \phi} 
\es \2 (\partial^\mu \varphi )^2 
     +
     \2 
     g'^2 (A^\mu + \partial^\mu \theta )^2  \varphi^2 \ .
\eeq
The modulus field can be written as $\varphi = v + h$, 
where $v$ will be treated as a parameter of the FF 
theory and the field $h$ may vary in space-time. When 
one sets $h=0$, the potential $\cV_\phi$ in Eq.~(\ref{Lphi}) 
has its minimal value $-\mu^4/(2\lambda^2)$ for $v = 
\sqrt{2} \ \mu/ \lambda$. Using this special value of $v$, 
one has
\beq
\cV(\phi)
\rs
- { \mu^4 \over 2 \lambda^2} + \2 \, (\sqrt{2} \, \mu )^2 \ h^2+ { \lambda \over \sqrt{2} } \ \mu \ h^3 + {\lambda^2 \over 8} \
h^4 \ .
\eeq
The Lagrangian density of Eq.~(\ref{cL}) is invariant under substitutions
\beq
\label{gaugepsi}
\psi       \es   e^{-ig f} \tilde \psi  \ , \\
\label{gaugeA}
A^\mu   \es  \tilde A^\mu + \partial^\mu f \ , \\
\label{gaugevarphi}
\varphi  \es \tilde \varphi \ , \\
\label{gaugetheta}
\theta   \es \tilde \theta  - f  \ .
\eeq
The meaning of this invariance is that the Lagrangian 
density is the same function of fields with and without 
the tilde. The corresponding minimal coupling that 
appears in a quantum theory obtained using the RGPEP, 
will be discussed in Sec.~\ref{firstorderterms}, see
Eqs.~(\ref{HpsiApsi2}) and (\ref{HpsiBpsi2}).

\subsection{ Massive limit }
\label{massivelimit}

Consider the limit of $g' \to 0$, $v \to \infty$ and 
$g' v = \kappa$ kept constant, which will be called 
the {\it massive} limit. In this limit, 
\beq
\label{Aphimassive}
\cL_{A \phi} 
\es \2 ( \partial^\mu h )^2 
     +
     \2 
     \kappa^2 (A^\mu + \partial^\mu \theta )^2 \ , \\
\label{Lphi2massive}
\cV_\phi
\es
- { \mu^4 \over 2 \lambda^2} 
+ \2 \, (\sqrt{2} \, \mu )^2 \ h^2 
\ .
\eeq
The field $h$ decouples and retains an arbitrary mass 
$\sqrt{2} \ \mu$. 

\subsection{ Gauge choice $f=-\theta$ }
\label{gaugef=-theta}

Using $f=-\theta$ one obtains 
\beq
\label{Lpsitheta1}
\cL_\psi 
\es \bar {\tilde \psi} \left[ \left( i \partial_\mu - g \tilde
A_\mu \right)
\gamma^\mu - m \right] \tilde \psi \ , \\\label{LA3theta1}
\cL_A 
\es - {1 \over 4} \tilde F_{\mu \nu} \tilde F^{\mu \nu} \ , \\
\label{Aphi3theta1}
\cL_{A \phi} 
\es 
     \2 (\partial^\mu \tilde \varphi )^2 
     +
     \2 
     g'^2 \tilde A^{\mu \, 2}  \tilde \varphi^2
     \ , \\
\label{Lphi3theta1}
\cV_\phi
\es
\cV ( \tilde \varphi/\sqrt{2}) \ .
\eeq
In the massive limit, $\cL_{A\phi}$ is
\beq
\cL_{A \phi} 
\es \2 (\partial^\mu \tilde h)^2 
     +
     \2 
     \kappa^2 \tilde A^2   \ ,
\eeq
the potential $\cV_\phi$ reduces to $\mu^2 \tilde h^2$ plus a
constant $\mu^2/(2 \lambda^2)$ that can be ignored,
while the densities $\cL_\psi$ and $\cL_A$ remain unchanged. 
The resulting action corresponds to a free scalar field $\tilde h$ of
mass $\sqrt{2} \, \mu$ and a vector field $\tilde A$ of mass $\kappa$
minimally coupled to the fermion field $\tilde \psi$. The massive-limit 
theory with field $\tilde h$ removed turns out to be the same as 
Soper's~\cite{Soper} when one identifies his field $B$ with 
our $-\kappa \theta$ and his mass parameter $\kappa$ with 
our $\kappa = g'v$. 

If the gauge symmetry under consideration were realized in nature 
and photons indeed had a very small mass~$\kappa$, which is 
theoretically possible~\cite{GoldhaberNieto}, there would also exist 
a decoupled scalar field $h$ of unknown mass, as the FF of the 
theory in the massive limit indicates, too. According to~\cite{PDG}, 
the photon mass is smaller than $10^{-18}$eV/$c^2$. Searches 
for new forms of matter are motivated by data concerning the 
structure and evolution of the universe, besides questions concerning 
the standard model.

\section{ Canonical FF Hamiltonian }
\label{canonicalFFhamiltonian}

In the FF of dynamics, the space-time coordinate 
$x^+ = x^0+x^3$ is used as the evolution 
parameter analogous to time in the instant form (IF)~\cite{DiracFF}.
The coordinates $x^-=x^0-x^3$ and $x^\perp = (x^1,x^2)$ 
parameterize points on the space-time hyperplanes that are 
defined by fixed values of $x^+$. These hyperplanes are called 
``light fronts'' or just ``fronts,'' such as the front defined by the 
condition $x^+=0$. Evolution in $x^+$ from the front at
$x^+=0$ to other fronts is generated by the Hamiltonian $P^-$.

Field theory relates the Lagrangian density of Eq.~(\ref{cL}) to the
corresponding Hamiltonian density through the energy-momentum 
tensor density $T^{\mu \nu}$, 
\beq
\label{Tmunu}
T^{\mu \nu} 
\es
\sum_\chi
       { \partial \cL \over \partial \partial_\mu \chi } \
       \partial^\nu \chi
-
g^{\mu \nu}  \cL  \ ,
\eeq
where $\chi$ stands for a field in a theory. 
The FF Hamiltonian density is $\cH = T^{+ \, -}/2$ 
and the Hamiltonian $P^-$ is given by~\cite{KogutSoper,Yan3}
\beq
\label{Pminus}
P^- \es \int d^2 x^\perp dx^- \  \cH \ .
\eeq
where the integral extends over the front at $x^+=0$. The 
Lagrangian density of Eq.~(\ref{cL}) is linear in $\partial^-f$ 
and the Hamiltonian  density is $\cH = - \cL(\partial^-f \to 0)$. 
For constructing a quantum theory, one needs to evaluate 
$\cH$ in terms of the fields' independent  degrees of freedom.

\subsection{  Equations of motion and gauge $\tilde A^+=0$ }
\label{EMandgaugeA+=0}

The principle of minimal action with the Lagrangian desity 
of Eq.~(\ref{cL}) implies the Euler-Lagrange (EL) equations 
that, when written in terms of the fields 
$\psi$, $A$, $\varphi$ and $B=-\kappa \theta$, read
\beq
\label{ELpsiB}
\left[ \left( i \partial_\mu - g A_\mu \right)
\gamma^\mu - m \right] \psi \es 0 \ , \\
\label{ELAB}
\Box A^\beta - \partial^\beta \,  \partial_\alpha A^\alpha  
\es 
g \bar \psi \gamma^\beta \psi 
-
g'^2 \ \varphi^2 \ ( A^\beta - \kappa^{-1} \partial^\beta B ) \
, \\
\label{ELvarphiB}
\Box \varphi 
\es
g'^2 \ \varphi  \ ( A^\beta - \kappa^{-1} \partial^\beta B )^2 
- {\partial \cV(\varphi/\sqrt{2}) \over \partial \varphi} \ , \\\label{ELthetaB}
\partial_\mu
 g'^2 \varphi^2  (A^\mu - \kappa^{-1}  \partial^\mu B)  
\es 0 \ .
\eeq
The last equation is necessarily satisfied if the first 
two are. The first equation can be written in terms of 
the fermion field arranged according to the formula
$\psi = \psi_+ + \psi_-$, where $\psi_\pm = 
\Lambda_\pm \psi$ and $\Lambda_\pm = \2 \gamma^0 
\gamma^\pm = \2 ( 1 \pm \alpha^3)$ are $4 \times 4$ 
projection matrices. In these terms, the fermion EL 
equation is equivalent to two coupled equations,
\beq
\label{psi1}
(i \partial^- -g A^-) \psi_+  
- \left[ ( i\partial^\perp-gA^\perp) \alpha^\perp + m\beta
\right] \psi_- \es 0 \ , \\
\label{psi2}
(i \partial^+ - g A^+) \psi_- 
- \left[ ( i\partial^\perp-gA^\perp) \alpha^\perp + m\beta
\right] \psi_+ \es 0 \ .
\eeq
Using gauge symmetry, one can transform the fields
$\psi$, $A$, $\varphi$ and $B$ to $\tilde \psi$, $\tilde A$, 
$\tilde \varphi$ and $\tilde B$. The two coupled fermion 
equations have the same form in terms of the fields with
tilde and without tilde. However, if the gauge transformation
sets the field $\tilde A^+$ to zero, then 
\beq
\label{psi3}
\tilde \psi_-
\es
{1 \over i \partial^+ } 
\left[ ( i\partial^\perp-g \tilde A^\perp) \alpha^\perp + m\beta \right] \tilde \psi_+ \ .
\eeq
The field $\tilde \psi_-$ on a front is thus given in terms of 
the fields $\tilde \psi_+$ and $\tilde A^\perp$ on the same 
front. Similarly, the EL Eq.~(\ref{ELAB}) for $\beta = +$ in 
the gauge $\tilde A^+=0$ constrains the field $\tilde A^-$, 
\beq
\label{constraintA}
\tilde A^- \es { 2 \over \partial^+ } \partial^\perp \tilde A^\perp
-
{2 \over \partial^{+ \, 2} }
\left( g \bar {\tilde \psi} \gamma^+ \tilde \psi 
+
g'^2 \ \tilde \varphi^2 \  \kappa^{-1} \partial^+ \tilde B  
\right) \ .
\eeq
As a consequence of the constraints, the FF Hamiltonian 
density is a function of fields $\tilde \psi_+$, $\tilde A^\perp$, 
$\tilde B$ and $\tilde \varphi$.

\subsection{ Hamiltonian density $\cH$ }
\label{hamiltoniandensitycH}

We use the Lagrangian density $\cL$ of Eq.~(\ref{cL}) written 
in terms of the independent field degrees of freedom $\tilde \psi_+$, 
$\tilde A^\perp$, $\tilde B$ and $\tilde \varphi$, to evaluate 
the Hamiltonian density using Eq.~(\ref{Tmunu}) for $T^{+-}= 2 \cH$.
From now on, we omit the tilde and employ notation $\varphi = v +h$
and $\kappa = g'v$. We also introduce the fields $\psi_f$ and $A_f$ 
that are given by the constraint Eqs.~(\ref{psi3}) and (\ref{constraintA}) 
in the absence of interaction~\cite{LepageBrodsky},
\beq
\label{Aminus1}
A_f^- \es {2 \over \partial^+} \partial^\perp A^\perp \ ,
\quad
A_f^+       \rs 0 \ , 
\quad
A_f^\perp  \rs A^\perp \ , \\
\psi_{f+} \es \psi_+ \ , 
\quad
\psi_{f-} \rs {1 \over i \partial^+} \
\left[ \alpha^\perp \, i\partial^\perp + m\beta \right] \psi_+ \ .
\eeq
The Hamiltonian density reads
\beq
\label{cH14}
\cH \es 
\2 \left\{ 
{1 \over \partial^+ }
\left[ g \bar \psi \gamma^+ \psi - 2 \kappa B \, (1+h/v) \,
\partial^+h/v \right] \right\}^2
\np
(1+h/v)^2 \kappa B \
{1 \over \partial^+ }
\left[ g \bar \psi \gamma^+ \psi - 2 \kappa B \, (1+h/v) \,
\partial^+h/v \right]
\np
\bar \psi_f \2 \gamma^+ { (i \partial^\perp)^2 + m^2 \over
i\partial^+} \psi_f
+
g \bar \psi_f \not \hspace{-4pt} A_f \psi_f 
+
\2 g^2 \bar \psi_f \not \hspace{-4pt} A_f { \gamma^+ \over i
\partial^+} \not \hspace{-4pt} A_f \psi_f
\nm
\2 A_f^\mu \left[ (i\partial^\perp)^2 + \kappa^2 (1 + h/v)^2
\right] A_{f \mu}
+
(1+h/v)^2 \kappa \ A_f^\mu \ \partial_\mu B
\np \2 h \left[ (i\partial^\perp)^2 + (\sqrt{2} \, \mu )^2
\right] h
+  {\mu^2 \over v} h^3 + \left({\mu \over 2v} \right)^2 h^4 
-   (\mu v/2)^2 
\np 
\2 (1 + h/v)^2 \ B \left[ (i \partial^\perp )^2 + \kappa^2
(1+h/v)^2 \right] B
-
(1+h/v) B \ \partial^\perp B \ \partial^\perp h/v \ .
\eeq
It differs from Soper's, because it involves additional fields. 
However, in the massive limit that ignores quantum effects, 
see Sec.~\ref{massivelimit}, in which $ g' \to 0$,  $v \to 
\infty$, $g'v = \kappa$ is kept constant (we could also 
consider the additional limit $\mu v \to 0$ to eliminate the 
constant $-(\mu v/2)^2$ and hence arrive at massless $h$), 
one obtains
\beq
\label{cH14}
\cH \tl
\2 \left[ {1 \over \partial^+ } g \bar \psi \gamma^+ \psi
\right]^2
+
\kappa B \ {1 \over \partial^+ } \ g \bar \psi \gamma^+ \psi  
\np
\bar \psi_f \2 \gamma^+ { (i \partial^\perp)^2 + m^2 \over
i\partial^+} \psi_f
+
g \bar \psi_f \not \hspace{-4pt} A_f \psi_f 
+
\2 g^2 \bar \psi_f \not \hspace{-4pt} A_f { \gamma^+ \over i
\partial^+} \not \hspace{-4pt} A_f \psi_f
\np
\2 A_f^i \left[ (i\partial^\perp)^2 + \kappa^2 \right] A_f^i 
+
\kappa \ A_f^\mu \ \partial_\mu B
\np \2 h \left[ (i\partial^\perp)^2 + (\sqrt{2} \, \mu)^2 \right] h 
\np 
\2 \ B \left[ (i \partial^\perp )^2 +  \kappa^2 \right] B \ .
\eeq
The second term, with the field $B$ and fermion plus current, can 
be replaced by the one that is equivalent through integration by parts. 
Since $A_f^+=0$ and $\partial_\mu A_f^\mu=0$, the seventh term 
that couples field $A_f$ to the gradient of filed $B$ is equivalent to 
zero. The decoupled field $h$ will be ignored in further discussion. 
Thus, one obtains the Hamiltonian density that is precisely equivalent 
to Soper's for massive QED~\cite{Soper}. It can be written as 
\beq
\label{cH16}
\cH \es
\bar \psi_f \gamma^+ { (i \partial^\perp)^2 + m^2 \over 2
i\partial^+} \psi_f
+
\2 A_f^i \left[ (i\partial^\perp)^2 + \kappa^2 \right] A_f^i 
+
\2 \ B \left[ (i \partial^\perp )^2 +  \kappa^2 \right] B 
\np
g \bar \psi_f \not \hspace{-4pt} A_f \psi_f 
-
g \bar \psi_f \gamma^+ \psi_f  {\kappa \over i\partial^+ } iB
+
\2 g^2 \bar \psi_f \not \hspace{-4pt} A_f { \gamma^+ \over i
\partial^+} \not \hspace{-4pt} A_f \psi_f
+
\2 \left[ {1 \over \partial^+ } g \bar \psi_f \gamma^+ \psi_f
\right]^2 \ .
\eeq
If the coupling constant $g$ were set to zero, the first 
three terms would describe the free fermion field $\psi_f$, 
free gauge boson field $A_f$ with two polarizations
and a free scalar field $B$. The fourth and fifth terms 
describe the minimal coupling of fields $A_f$ and $B$  
with fermions, respectively. The sixth term additionally 
couples transverse bosons to fermions as a result of 
the constraint Eq.~(\ref{psi3}). The last term is the FF 
fermion quartic interaction that results from the 
constraint Eq.~(\ref{constraintA}). It is a FF analog of 
the Coulomb term in the IF dynamics with its Gauss 
law. The Hamiltonian density of Eq.~(\ref{cH16}) is 
taken as a starting point for the canonical construction 
of a quantum theory {\it a la} Refs.~\cite{Soper,KogutSoper,Yan3}.

\subsection{ Quantization }
\label{quantization}

The quantum theory is introduced by replacing the fields 
$\psi_f$, $A_f$ and $B$ in Eq.~(\ref{cH16}) by the 
corresponding field operators on the front at $x^+=0$,
\beq
\label{hatpsi}
\hat \psi_f
\es
\sum_{\sigma = 1}^2  \int [p]
\left[  u_{p\sigma} \hat b_{p\sigma} e^{-ipx}
      +    v_{p\sigma} \hat d^\dagger_{p\sigma} e^{ipx}
\right]_{x^+=0}
\ ,
\\
\label{hatA}
\hat A_f^\mu
\es
\sum_{\sigma =1}^2  \int [p]
\left[  \varepsilon^\mu_{p\sigma} \hat a_{p\sigma} e^{-ipx}
+ \varepsilon^{\mu *}_{p\sigma} \hat a^\dagger_{p\sigma} e^{ipx}
\right]_{x^+=0}
\ ,
\\
\label{hatB}
\hat B
\es
\int [p]
\left[  -i \hat a_{p3} e^{-ipx}
        +  i \hat a^\dagger_{p3} e^{ipx}
\right]_{x^+=0} \ , 
\eeq
where $[p] = dp^+ \theta(p^+) d^2p^\perp /[2 p^+ (2\pi)^3]$.
Further, $u_{p\sigma}$ and $v_{p\sigma}$ are spinors for fermions 
of mass $m$~\cite{LepageBrodsky,fermions}. Symbols 
$\varepsilon_{p\sigma}$ denote polarization four-vectors for 
bosons~\cite{Soper,AbelianAPPB}. Thus, $\sigma$ labels 
fermions and gauge bosons that at rest have spin projections 
$\pm \2$ or $\pm 1$ on the $z$-axis, respectively. Further 
details of the notation are explained in App.~\ref{notation}. 
The creation and annihilation operators, denoted by $b$, $d$ 
and $a$, obey commutation or, in the case of fermions, 
anti-commutation relations of the form
\beq
[ \hat a_{p\lambda}, \hat a^\dagger_{q\sigma}]
\es
2 p^+ (2\pi)^3 \delta(p^+-q^+) \delta^2(p^\perp - q^\perp)
\delta_{\lambda \sigma} \ ,
\eeq
with other commutators or anti-commutators equal zero. 
The Hamiltonian $\hat P^-$ is obtained by integrating
the quantum density $\hat \cH$ on the front $x^+=0$
and normal ordering. 

At this point it is important to mention, on the basis of
hindsight, that the operators creating or annihilating quanta 
with infinitesimal $p^+$, {\it i.e.}, $p^+$ negligible in 
comparison with mass parameters $m$ and $\kappa$, 
including the case of $\kappa/m \to 0$, could contribute 
divergences to the free invariant masses of all physical 
states. Therefore, in the regulated and subsequently 
renormalized theory such quanta need to be suppressed. 
Formally, at this point one could introduce in Eqs.~(\ref{hatpsi}),
(\ref{hatA}) and (\ref{hatB}) an infinitesimal cutoff parameter 
$\epsilon^+$, imposing a condition $p^+ > \epsilon^+$ 
instead of $p^+ >0$. However, it will become self-evident 
in the next sections that, in the RGPEP, perturbatively calculated 
effective Hamiltonians for finite-size quanta with finite plus 
momenta are not sensitive at all to the cutoff parameter 
$\epsilon^+ \to 0$. Namely, it is shown in the next sections 
that the gauge boson mass $\kappa$ provides the required 
suppression through the vertex form factors that result 
from solving the RGPEP evolution Eq.~(\ref{RGPEP}). 
Regarding the divergent constants and one-particle operators 
that result from the normal ordering, they are dropped 
because constants do not count in the quantum dynamics 
and one-particle operators require counter terms anyway.
In summary, the cutoff on $p^+$ and normal ordering do 
not influence the content of a theory defined using the RGPEP.

\subsection{ Quantum Hamiltonian }
\label{quantumhamiltonian}
 
Our initial quantum Hamiltonian $\hat P^-$ is denoted 
by $\hat H$,
\beq
\label{initialcH}
\hat H \es \hat H_{\psi^2} + \hat H_{A^2} + \hat H_{B^2} + \hat H_{\psi A \psi} 
            + \hat H_{\psi B \psi} + \hat H_{\psi A A \psi} + \hat H_{(\psi\psi)^2} \ .
\eeq
The seven operators appear in one-to-one correspondence to
the seven terms in Eq.~(\ref{cH16}). To simplify notation for 
the quantum theory, the operator symbol ~$\hat{}$~ is omitted 
in further formulas. The first three terms are separately denoted by
\beq
\label{Hf}
H_f \es H_{\psi^2} + H_{A^2} + H_{B^2} \ ,
\eeq
where the subscript originates in the word {\it free}. The 
remaining four terms are denoted by $H_I$. All terms 
are given in full detail in App.~\ref{hamiltonianappendix}. 

\section{ Application of the RGPEP }
\label{ARGPEP}

The FF Hamiltonian of Eq.~(\ref{initialcH}) leads to divergences
and as such is not acceptable. The divergences can be identified 
and removed from the Hamiltonian using the RGPEP. We apply
it here in expansion in powers of the coupling constant $g$ up 
to and including terms order $g^2$. General introduction to the 
RGPEP and perturbative formulas for interactions of effective 
particles up to fourth order are available in~\cite{RGPEP}.   

In brief, the Hamiltonian $H$ of Eq.~(\ref{initialcH}) is used as 
an initial condition, $\cH_{t =0}=H$, for solving the differential equation 
\beq 
\cH'_t \es \left[ [ \cH_f, \tilde \cH_t ] , \cH_t \right] \ ,
\label{RGPEP}
\eeq 
where prime denotes differentiation with respect to the scale 
parameter $t=s^4$. The parameter $s$ has an intuitive 
interpretation of the size of effective quanta, see below. 
The tilde in $\tilde \cH_t$ indicates that each term in $\cH_t$ 
is multiplied by the square of total plus momentum carried by 
quanta annihilated or, equivalently, created by that term. Such
multiplication secures that Eq.~(\ref{RGPEP}) preserves
all kinematic symmetries of the FF of dynamics~\cite{DiracFF}. 
The double commutator used in Eq.~(\ref{RGPEP}) is introduced, 
following Wegner~\cite{Wegner}, to satisfy the requirement 
that the creation and annihilation operators for effective 
quanta of size $s$, denoted by $q_t$, are related to the 
initial ones, denoted by $q_0$, by such a unitary transformation 
$\cU_t$, 
\beq
\label{Uqt}
q_t \es \cU_t q_0 \cU_t^\dagger \ , 
\eeq 
that the Hamiltonian $\cH_t$ can only cause limited changes 
of the interacting quanta total invariant mass. The idea of 
replacing the Wilsonian principle of integrating out high-energy 
modes by the principle of integrating out large changes of 
energy dates back to Ref.~\cite{GlazekWilson}, which introduced 
the so-called similarity renormalization group procedure (SRG). 
The initial application of SRG to the FF Hamiltonian of QCD, 
using $P^-$ instead of energy, is outlined in Ref.~\cite{Wilsonetal}.
The RGPEP provides a relativistic extension of the latter idea. 
Instead of changes of $P^-$, we use changes of the invariant 
mass. Hence the motion of field quanta is not limited in any 
other way than by the speed of light. Also, instead of considering 
scale evolution of Hamiltonian matrices, the RGPEP uses operators. 
The number of quanta is not limited. These features are prerequisite 
for a complete formulation 
of a finite theory that includes the parton picture~\cite{FeynmanPM} 
of bound states as well as their spectroscopy.

The operator $\cH_t$ is defined to be a polynomial in 
the creation and annihilation operators that appear in 
Eqs.~(\ref{hatpsi}), (\ref{hatA}) and (\ref{hatB}). 
Solutions for the polynomial coefficients as functions of 
$t$ are found on the basis of their initial values in 
$H = \cH_{t=0}$. However, one has to remove 
divergences from the solutions. Therefore, the RGPEP
includes the alteration of the initial condition of $H = 
\cH_{t=0}$ by inclusion of additional terms that counter 
the divergences in solutions. In general, the counter
terms can only be found by successive approximations.
Solutions described in this exploratory article are limited 
to the lowest non-trivial order of series expansion in 
powers of the coupling constant.

To be more specific, solutions for the coefficients $c_t$ of order 
$g$ are of the form $c_t = f_{t \, 1} \,  c_0$, where $f_{t \, 1}$ 
is a unique form factor that vanishes exponentially fast when 
the difference between a total invariant mass of quanta created 
and a total invariant mass of quanta annihilated by the associated 
product of creation and annihilation operators exceeds $s^{-1}$. 
When imagined in terms of a matrix in the space of quantum 
states of specified total invariant mass (according to $\cH_f$), 
the Hamiltonian $\cH_t$ would appear band diagonal with the 
band width $\sim s^{-1}$. Now consider the second order. One 
obtains solutions of the generic form $f_{t \, 2} \, c_0^2$, since 
the initial Hamiltonian is squared. In a local theory, the intermediate 
states in the square of the Hamiltonian may have arbitrarily large 
invariant masses. Therefore, $c_0^2$ diverges when one sums 
over all the intermediate states. One has to regulate $c_0$ 
somehow to limit the sum and obtain finite $c_0^2$. So, $\cH_{t=0}$ 
is supplied with some regularization, which we denote by $r$. It 
is shown below how we do it for the Abelian gauge theory. To 
remove dependence of $\cH_t$ with finite $t$ on the regularization 
$r$, we need to include in $\cH_0$ a counter term $CT_{r2}$ of order 
$g^2$. Expansion to higher orders exhibits the same pattern. 
In addition, the actual expansion needs to be carried out using an 
effective coupling constant $g_t$~\cite{Gomez} instead of the 
initial $g$. However, the coupling constants $g_t$ and $g$ begin 
to differ first in third order calculation. In the present article only 
terms order 1, $g_t$ and $g_t^2$ are considered. Therefore, 
there is no need to distinguish $g_t$ from $g$ and we omit 
the subscript $t$ in $g_t$.
 
When one includes regularization factors $r$ and the 
corresponding counter terms $CT_r$, the initial Hamiltonian 
$\cH_0 = H$ of Eq.~(\ref{initialcH}) is changed to $H_r$,
\beq
\label{initialcHr}
H_r \es H_f + H_{\psi A \psi \, r} 
                 + H_{\psi B \psi \, r} + H_{\psi A A \psi \, r} + H_{(\psi\psi)^2 \, r} 
                 + CT_r \ .
\eeq
Thus the initial Hamiltonian $\cH_0$ takes the form of 
a computable series in powers of the coupling constant 
\beq
\label{Hr1}
H_r
\es 
H_f + g H_{r1} + g^2 H_{r2} + g^2 CT_{r2}  + O(g^3) \  .
\eeq
Correspondingly, solution of Eq.~(\ref{RGPEP}) also has 
the form of a series
\beq
\label{cHtexp}
\cH_t
\es
\cH_f + g \cH_{t1} + g^2 \cH_{t2} + O(g_t^3) \ .
\eeq
To calculate the terms in this series one equates coefficients 
of the same powers of $g$ on both sides of Eq.~(\ref{RGPEP}) 
and obtains equations
\beq
\label{RGPEP0}
\cH_f' \es 0 \ , \\
\label{RGPEP1}
\cH_{t\,1}' \es \left[ [\cH_f, \tilde \cH_{t\,1}], H_f \right] \ ,
\\
\label{RGPEP2}
\cH_{t\,2}' \es \left[ [\cH_f, \tilde \cH_{t\,2}], H_f \right] +
\left[ [\cH_f, \tilde \cH_{t\,1}], \cH_{t\,1} \right] \ .
\eeq
These are solved in the following sections. In the last
step of solving for the renormalized Hamiltonians $H_t$,
the canonical operators $q_0$ are replaced by the effective 
ones, $q_t$, according to the formula $H_t = \cH_t(q_0 
\to q_t)$. To simplify our notation below, the operators 
$q_0$ are denoted by $q$, {\it i.e.}, the subscript 0 is 
omitted. Thus,
\beq
H_t \es \cU_t \cH_t \cU_t^\dagger \rs \cH_t(q \to q_t) \ .
\eeq
The perturbative expansion for $\cH_t$ 
in Eq.~(\ref{cHtexp}) directly implies a 
similar one for $H_t$,
\beq
\label{Htexp}
H_t
\es
H_{t f} + g H_{t1} + g^2 H_{t2} + O(g_t^3) \ .
\eeq
The discussion that follows is mostly
carried out in terms of the operator $\cH_t$. 

\subsection{ Free Hamiltonian terms }
\label{Hf}

Since the free Hamiltonian $\cH_f$ obeys $\cH_f'=0$, 
see Eq.~(\ref{RGPEP0}), it is given by the canonical 
Eqs.~(\ref{Hpsi2app}), (\ref{HA2app}) and (\ref{HB2app})
in App.~\ref{hamiltonianappendix}. To obtain $H_{t f}$, 
the creation and annihilation operators $q_0$ for bare, 
point-like quanta are replaced in $\cH_f$ by the operators 
$q_t$ for effective particles of size $s$, with the same 
quantum numbers. So, 
\beq
\label{Htf}
H_{t f} \es H_{t \, \psi^2} + H_{t \, A^2} + H_{t \, B^2} \ ,
\eeq
where
\beq
\label{Hpsi2}
H_{t \, \psi^2} 
\es 
\sum_{\sigma = 1}^2 \int [p] \ {p^{\perp \, 2} + m^2 \over p^+}
\
  \left[b^\dagger_{t \, p\sigma }b_{t \, p\sigma } + 
  d^\dagger_{t \, p\sigma }d_{t \, p\sigma } \right] \ , \\
\label{HA2}
H_{t \, A^2} 
\es 
\sum_{\sigma =1}^2 \int [p] \ {p^{\perp \, 2} + \kappa^2 \over
p^+} \
  a^\dagger_{t \, p\sigma}a_{t \, p\sigma} \ , \\
\label{HB2}
H_{t \, B^2} 
\es 
\int [p] \ {p^{\perp \, 2} + \kappa^2 \over p^+} \
  c^\dagger_{t \, p} c_{t \, p} \ .
\eeq

\subsection{ First-order interaction terms }
\label{firstorderterms}

According to Eq.~(\ref{RGPEP1}), 
the coefficients $h_{t \, 1 \, ca}$ of products $c$ and $a$ of 
creation and annihilation operators, respectively, under the 
momentum integrals in $\cH_{t \, 1}$, satisfy the differential 
equations
\beq
h'_{t\,1\,ca} \es - ( \cM_c^2 - \cM_a^2)^2 \ h_{t \, 1 \, ca} \ ,
\eeq
where $\cM_a$ denotes the invariant mass of particles
annihilated and $\cM_c$ particles created by the interaction.
The initial conditions at $t=0$, denoted by $h_{0 \, 1 \, ca}$, 
are provided by the first-order canonical coefficients 
shown in Eqs.~(\ref{HpsiApsi}) and (\ref{HpsiBpsi}) 
in App.~\ref{hamiltonianappendix}. Thus, solutions 
for the coefficients are
\beq
h_{t\,1\,ca} \es f_{t \, c.a} \ h_{0 \, 1\,ca} \ ,
\eeq
which amount to the initial conditions multiplied   
by the RGPEP vertex form factors,
\beq
\label{ftc.a}
f_{t \, c.a}
\es
\exp{ [ - t ( \cM_c^2 - \cM_a^2)^2 ] } \ .
\eeq
The RGPEP form factors suppress the invariant mass changes 
that exceed $1/s$ exponentially fast. The suppression allows 
one to intuitively associate the parameter $s$ with the concept 
of size of effective quanta. The local gauge theory corresponds 
to point-like quanta and $s=0$. The larger $s$ the stronger 
the vertex suppression. Large $s$ implies that only small
changes of the off-shell departures of virtual interacting quanta 
can occur. This correlation is similar to the one found in quantum 
mechanics of bound states of charged particles, whose form 
factors suppress absorption or emission of light with momentum 
that exceeds the inverse of their size. However, one should keep
in mind that the RGPEP effective quanta can be in arbitrary 
relativistic motion with respect to each other and they do not 
behave as bound states known in non-relativistic quantum 
mechanics, so that the interpretation of $s$ as quantum-mechanical 
size is merely based on an analogy. 

The RGPEP form factor $f_{t \, c.a}$ appears in front of all 
products of creation and annihilation operators in every 
interaction term equally. Namely,
\beq
\label{HpsiApsit}
H_{t \, \psi A \psi} 
\es   
g \sum_{123}\int[123] \,
  \tilde \delta_{c.a} \, f_{t+t_r \, c.a} \nt
  \left[
\bar u_2 \hspace{-3pt} \not\!\varepsilon_1^* u_3 \
b^\dagger_{t\,2} a^\dagger_{t\,1} b_{t\,3}
- \bar v_3 \hspace{-3pt} \not\!\varepsilon_1^* v_2 \
d^\dagger_{t\,2} a^\dagger_{t\,1} d_{t\,3}
+\bar u_1 \hspace{-3pt} \not\!\varepsilon_3 v_2 \
b^\dagger_{t\,1} d^\dagger_{t\,2} a_{t\,3} \ + \ h.c.
  \right] 
\ , \\
\label{HpsiBpsit}
H_{t \, \psi B \psi}
\es
- g \sum_{23}\int[123] \,
  \tilde \delta_{c.a} \, f_{t+t_r \, c.a} \nt
  \left[
\bar u_2 { \kappa \gamma^+ \over p_1^+} u_3 \ b^\dagger_{t\,2}
c^\dagger_{t\,1} b_{t\,3}
- \bar v_3 { \kappa \gamma^+ \over p_1^+} v_2 \ d^\dagger_{t\,2}
c^\dagger_{t\,1} d_{t\,3}
+\bar u_1 { \kappa \gamma^+ \over p_3^+} v_2 \ b^\dagger_{t\,1}
d^\dagger_{t\,2} c_{t\,3} \ + \ h.c.
  \right]          \ ,
\eeq
Note that the canonical creation and annihilation operators for
initial, point-like quanta are replaced by the operators for quanta 
of size $s$, corresponding to $t=s^4$. 

The operator structure of the first-order solutions resembles 
the canonical one, so that for momenta for which $f_{t+t_r \, c.a} 
\sim 1$, one has
\beq
\label{HpsiApsi2}
H_{t \, \psi_t A_t \psi_t} 
\es
H_{{\rm can} \, \psi_t A_t \psi_t}    
\ , \\
\label{HpsiBpsi2}
H_{t \, \psi_t B_t \psi_t}
\es
H_{{\rm can} \, \psi_t B_t \psi_t}
\ .
\eeq 
The subscript ``can'' refers to the canonical minimal coupling
Hamiltonian terms. Fields with subscript $t$ are built from 
creation and annihilation operators $q_t$ in the same way 
as the canonical quantum fields are built from the operators
$q_0$. The two Eqs.~(\ref{HpsiApsi2}) and (\ref{HpsiBpsi2}) 
express the RGPEP interpretation of gauge symmetry as a 
guiding principle in constructing relativistic quantum theory 
of particles : {\it The effective minimal coupling Hamiltonian 
interaction term appears for momentum transfers much smaller 
than $s^{-1}$ equal to the canonical minimal coupling term 
in a local gauge theory.} The difference that is hard to 
recognize is the one between the operators $q_t$ and $q_0$. 

The above interpretation implies also that the regularization 
factors introduced in Eq.~(\ref{initialcHr}) can be just the 
RGPEP vertex form factors $f_t$ with some extremely 
small value of $t$, denoted by $t_r$~\cite{AbelianAPPB}. 
Precisely this regularization is the origin of the sum $t+t_r$ 
as a size parameter in the vertex form factors displayed in 
the solutions of Eqs.~(\ref{HpsiApsit}) and (\ref{HpsiBpsit}). 
When $t \to 0$, the regularization parameter $t_r$ remains 
and makes the form factor regulate the Hamiltonian. The 
regularization is lifted when $t_r$ is sent to zero.

It is now visible that in the tree approximation the regularization 
influence on the renormalized theory vanishes when the regularization 
is lifted. Namely, for a fixed finite $t$ the infinitesimal $t_r$ is 
inconsequential. The regularization factors $f_{t_r\,c.a}$ are said 
to be {\it muted} as functions of momenta by the RGPEP vertex 
form factors $f_{t\,c.a}$ with finite $t$ when $t_r \to 0$. In general, 
the condition that the RGPEP factors mute regularization factors in 
a finite effective theory at the tree level implies that the gauge 
symmetry becomes manifest in the low-energy tree-level processes 
that involve momentum changes much smaller than the inverse size 
of the effective particles.

\subsection{ Fermion self interactions }
\label{fermionselfinteractions}

As a result of second-order self-interactions, the mass-squared 
terms for fermions change from $m^2$ in the canonical
Hamiltonian to $m^2 + g^2 \delta m^2(t)$  in $H_t$. The 
corrected mass appears in the coefficients of operators 
$b^\dagger_{t \, p\sigma } b_{t \, p\sigma }$ and 
$d^\dagger_{t \, p\sigma }d_{t \, p\sigma }$ in $H_t$. One 
calculates $\delta m^2(t)$ by integrating its derivative with 
respect to $t$ that is contained Eq.~(\ref{RGPEP}). For terms 
order $g^2$,  Eq.~(\ref{RGPEP}) reduces to Eq.~(\ref{RGPEP2}). 
The coefficients of operators $b^\dagger_{p\sigma } b_{p\sigma }$ 
and $d^\dagger_{p\sigma } d_{p\sigma }$ are extracted from 
the right-hand side of Eq.~(\ref{RGPEP2}). The only contributions 
come from the second term,
\beq
\cH_{t\, 2 \, \delta m^2}' 
\es 
\left[ [\cH_f, \tilde \cH_{t\,1}], \cH_{t\,1} \right]_{\delta m^2}
\ .
\eeq
More specifically, from creation and subsequent 
annihilation of a fermion and a boson. That set of quanta 
is symbolized by $f b$. Our calculation yields, in notation 
explained in App.~\ref{notation},
\beq
\label{pom3}
\cH_{t\, 2 \, \delta m^2}' 
\es 
 \sum_3 \int[3]
{(\delta m_2^2)' \over p_3^+} \ \left[ b^\dagger_3 b_3 +
d^\dagger_3 d_3 \right]
\ ,
\eeq
where
\beq
\label{deltamprime}
(\delta m_2^2)'
\es
  -2
  \int[xk] \ f_{t+t_r \, fb.f}^2 \ \left( \cM_{fb}^2 - m^2 \right) \ F_{\delta m^2} \ , \\
 F_{\delta m^2} \es 2
 \left[ {k^{\perp\,2} + x^2 m^2 \over 1-x} 
 +  
 2 { k^{\perp\,2}  +  (1-x) \kappa^2 \over x^2 } \right] \ , \\
 f_{t+t_r \, fb.f}^2 
 \es
 e^{ - 2 (t+t_r) (\cM_{fb}^2-m^2)^2} \ , \\
\cM_{fb}^2 \es { k^{\perp\,2}  +  m^2 \over 1-x }+ { k^{\perp\,2}  +  \kappa^2 \over x } \ .
 \eeq
The variables $x$ and $k^\perp$ denote components
of the boson momentum in the fermion self-interaction
set $f b$. In evaluation of the factor $F_{\delta m^2}$, 
contributions of quanta of field $B$ turn out to amount 
to just adding $\kappa^2$ to $p^{\perp 2}$ in the sum 
over polarizations of field-$A$ quanta with momentum 
$p$. Integration over $t$ in Eq.~(\ref{deltamprime}) 
results in
\beq
\delta m^2(t) 
\es
 \delta m^2(0)
 -
  \int[xk] \
 \left(  f_{t_r \, fb.f}^2   -   f_{t+t_r \, fb.f}^2  \right) 
  \
\left( \cM_{fb}^2 - m^2 \right)^{-1}
F_{\delta m^2} 
\ .
 \eeq
In the limit of $t_r$ going to zero that lifts the regularization, 
the integral diverges. The divergence can be canceled by 
adjusting the value of $\delta m^2(0)$.  However, the 
finite part of $\delta m^2(0)$ can only be fixed by
comparison of theory with data. 

Directly relevant observable is the Hamiltonian eigenvalue
$p^- = (p^{\perp 2}+m_f^2)/p^+$, in which $m_f$ stands
for the smallest mass eigenvalue for the eigenstates with 
fermion quantum numbers. In the present calculation, one 
considers the eigenstates approximated by a superposition 
of effective single fermion and two-body effective fermion-boson 
Fock states. The momentum components $p^+$ and $p^\perp$ 
are the eigenvalues of kinematic Poincar\'e generators of front 
translations, $\hat P^+$ and $\hat P^\perp$. These eigenvalues
drop out entirely from the fermion eigenvalue equation and the 
eigenvalue reduces to $m_f^2$. For $m_f^2$ to match $m^2$ in 
Eq.~(\ref{Hpsi2}) for arbitrary finite values of $t$, the 
counter term must be
\beq
\delta m^2(0)
\es
  \int[xk] \ f_{t_r \, fb.f}^2 \
\left( \cM_{fb}^2 - m^2 \right)^{-1}
F_{\delta m^2} 
\ .
\eeq
This condition determines the counter term including its 
finite part. The result for $\delta m^2(t)$ is 
\beq
\label{deltam2t}
\delta m^2(t) 
\es
\int[xk] \ f_{t + t_r \, fb.f}^2 \
\left( \cM_{fb}^2 - m^2 \right)^{-1}
F_{\delta m^2} 
\ , 
\eeq
where for any finite value of $t$ the limit of 
no regularization is obtained by letting $t_r$ 
tend to zero. As a result, the mass-squared 
Hamiltonian term for effective fermions of 
size $s = t^{1/4}$ is corrected by a term 
order $g^2$ of the form
\beq
\label{Hpsi2g2}
H_{t \, 2 \, \delta m^2} 
\es 
\sum_{\sigma = 1}^2 \int [p] \ { \delta m^2(t) \over p^+}
\
  \left[b^\dagger_{t \, p\sigma }b_{t \, p\sigma } + 
  d^\dagger_{t \, p\sigma }d_{t \, p\sigma } \right] \ .
\eeq
This term is included as a part of the entire Hamiltonian 
$H_t$. The latter is used to calculate masses of bound 
states of fermions. Plots that show how the function 
$\delta m^2(t)$ arises are provided in Sec.~\ref{plots}.

\subsection{ Boson self interactions }
\label{bosonselfinteractions}

Mass corrections for the effective gauge boson quanta
of size $s = t^{1/4}$ are determined according to 
the same algorithm as for the fermions. One integrates
their derivatives given in the RGPEP Eq.~(\ref{RGPEP}),
which in order $g^2$ reduces to Eq.~(\ref{RGPEP2}). 
Thus, the derivatives of the corrections order $g^2$ 
are obtained from
\beq
\cH_{t\, 2 \, \delta \kappa^2}' 
\es 
\left[ [\cH_f, \tilde \cH_{t\,1}], \cH_{t\,1} \right]_{\delta
\kappa^2} \ .
\eeq
One derives the derivatives of coefficients of terms $a^\dagger a$
and $c^\dagger c$ in $\cH_t$. The derivatives come from 
creation and subsequent annihilation of a fermion and an 
anti-fermion pair, symbolized by $f \bar f$. One integrates 
these derivatives from zero to $t$. The finite parts of counter terms 
in the initial condition at $t=0$ are defined by demanding that the 
mass-squared eigenvalues of $H_t$ for the gauge boson states 
are $\kappa^2$, equally for bosons of type $A$ and $B$. The 
resulting mass-squared terms for quanta of fields $A_t$ and 
$B_t$ turn out to differ from each other. Namely, we obtain
\beq
\label{Hdeltakappa2A}
H_{t \, 2 \, \delta \kappa_A^2} 
\es 
\sum_{\sigma =1}^2 \int [p] \ { \delta \kappa_A^2(t) \over
p^+} \
  a^\dagger_{t \, p\sigma}a_{t \, p\sigma} \ , \\
\label{Hdeltakappa2B}
H_{t \, 2 \, \delta \kappa_B^2} 
\es 
\int [p] \ { \delta \kappa_B^2(t) \over p^+} \
  c^\dagger_{t \, p} c_{t \, p} \ ,
\eeq
where
\beq
\label{kappaApom}
\delta \kappa_A^2(t) 
\es
  \int[xk] \ f_{t+t_r \, f \bar f.b}^2 \ \left( \cM_{f \bar f}^2 - \kappa^2 \right)^{-1}
  \ F_{\delta \kappa^2_A} \ , \\
\label{kappaBpom}
\delta \kappa_B^2(t) 
\es
  \int[xk] \ f_{t + t_r \, f \bar f.b}^2 \ \left( \cM_{f \bar f}^2 - \kappa^2 \right)^{-1}
   \ F_{\delta \kappa^2_B}  \ , 
\eeq
and
\beq
\cM_{f \bar f}^2 
\es { k^{\perp \, 2} + m^2 \over x (1-x) } \ , \\
F_{\delta \kappa^2_A} 
\es 
2 { [x^2 + (1-x)^2] k^{\perp \, 2} + m^2 \over x (1-x)  } \ , \\
F_{\delta \kappa^2_B} 
\es 8 \kappa^2 \, x(1-x) \ .
\eeq
The transverse gauge boson mass is corrected by a term 
that varies rapidly with $t$. The third-polarization gauge 
boson mass is proportional to $\kappa^2$ and does not 
exhibit any such rapid variation with $t$. Detailed discussion 
of how the functions $\delta \kappa_A^2(t)$ and $\delta 
\kappa_B^2(t)$ arise is postponed to Sec.~\ref{plots}. 

\subsection{ Boson exchange }
\label{bosonexchange}

For the purpose of discussion of an example of effective 
bound-state dynamics in Sec.~\ref{plots}, we consider 
the Hamiltonian interaction terms of order $g^2$ that 
involve exchanges of gauge bosons of types $A$ and
$B$ between a fermion and an anti-fermion. The 
RGPEP evolution of second-order interaction terms is 
obtained from Eq.~(\ref{RGPEP2}). We focus on the
coefficients $c(121'2')$ of operators $b_1^\dagger 
d_2^\dagger d_{2'} b_{1'}$. The fermions that come 
out of the interaction carry quantum numbers labeled 
by 1. The anti-fermions come out with quantum numbers 
labeled by 2. The fermions and anti-fermions that come 
in carry quantum numbers labeled by $1'$ and $2'$, 
correspondingly. When it is useful, we abbreviate notation 
for these coefficients or for the operators that contain them 
by using the acronym or subscript $q \bar q$, associating 
$q$ with fermions and $\bar q$ with anti-fermions, {\it 
a la} positronium or quarkonia. The purpose of using 
$q\bar q$ instead of $f\bar f$ in this section is that the 
subscript $f$ is more useful here to indicate the free 
part of the Hamiltonians and the RGPEP form factors,
instead of fermions. 

The boson-exchange terms are contained in Eq.~(\ref{RGPEP2}) of the form,
\beq
\label{h2prime1}
\cH_{t\,2 \, q \bar q}' \es  \left[ [\cH_f, \tilde \cH_{t\,2\, q \bar q}], \cH_f \right] 
+   \left[ [\cH_f, \tilde \cH_{t\,1}], \cH_{t\,1} \right]_{q \bar q} \ .
\eeq
The initial condition includes the regulated canonical
$q \bar q$ interaction term and, potentially, a counter 
term that needs to be calculated.  The initial-condition 
canonical term consists of
\beq
\label{hfbarf}
H_{q\bar q \, r \, \rm can}
\es
g^2 
\sum_{121'2'}\int[121'2'] \,
\tilde \delta_{12.1'2'} \, r_{121'2'} 
\ h_{0 \, {\rm can} \, 2 \, q \bar q}(121'2')
\
b_1^\dagger d_2^\dagger d_{2'} b_{1'}    
\ ,
\eeq 
where, on the basis of Eq.~(\ref{Hpsi2ordered}),
\beq 
h_{0 \, {\rm can} \, 2 \, q \bar q}(121'2')
\es
-
{ \bar u_1 \gamma^+ u_3 \ \bar v_4 \gamma^+ v_2 \over (p_1^+ -
p_3^+)^2 }
+
{ \bar u_1 \gamma^+ v_2 \ \bar v_4 \gamma^+ u_3 \over (p_1^+ +
p_2^+)^2 }
\ . 
\eeq
The first term corresponds to the FF instantaneous interaction
that is analogous to the Coulomb term in the IF Hamiltonian. The
second term corresponds to the FF instantaneous interaction 
through the annihilation channel rather than the exchange. We 
discuss the $q \bar q$ annihilation channel interaction along 
our discussion of the boson-exchange interaction since both can 
contribute to the dynamics of the $q \bar q$ bound states. Both 
interactions result from the FF constraint Eq.~(\ref{constraintA}) 
for $A^-$, analogous to the IF Gauss law. However, instead of 
the inverse of Laplacian they involve only the inverse of $\partial^{+2}$. 
The factor $r_{121'2'}$ provides regularization, according to the 
rules set at the end of Sec.~\ref{firstorderterms} and in 
App.~\ref{Appregularization}. Namely, the form factor $f_{t_r}$ 
with infinitesimal $t_r$ is inserted in both fermion currents 
that appear in the interaction. 

Following~\cite{RGPEP} and using notation defined in 
App.~\ref{notation}, integration of Eq.~(\ref{h2prime1}) 
begins with writing $\cH_{t \, 2 \, q\bar q}$ in the form 
\beq
\label{hfbarf}
\cH_{t \, 2 \, q\bar q}
\es
\sum_{121'2'}\int[121'2'] \,
\tilde \delta_{12.1'2'} 
\ h_{t \, 2 \ q \bar q}(121'2')
\
b_1^\dagger d_2^\dagger d_{2'} b_{1'}    
\ .
\eeq 
The differential equation to solve reads
\beq
\label{h2prime2}
&&
\sum_{121'2'}\int[121'2'] \,
\tilde \delta_{12.1'2'} 
\
h'_{t \, 2 \, q \bar q}(121'2')
\
b_1^\dagger d_2^\dagger d_{2'} b_{1'}    \\
\es
-\sum_{121'2'} \int[121'2'] \,
\tilde \delta_{12.1'2'} \
(\cM_{12}^2 - \cM_{1'2'}^2)^2 \ h_{t \, 2 \, q \bar q}(121'2')
\
b_1^\dagger d_2^\dagger d_{2'} b_{1'}  
\np
\left[ [\cH_f, \tilde \cH_{t\,1}], \cH_{t\,1} \right]_{q \bar q} \
.
\eeq
Writing
\beq
\label{h2-121'2'}
h_{t \, 2 \, q \bar q}(121'2')
\es
e^{-t (\cM_{12}^2 - \cM_{1'2'}^2)^2 } g_{t \, 2 \, q \bar q}(121'2') \
,
\eeq
one obtains a differential equation for $g_{t \, 2 \, q \bar q}(121'2')$,
\beq
\label{cGprime}
\sum_{121'2'}\int[121'2'] \,
\tilde \delta_{12.1'2'} \
e^{-t (\cM_{12}^2 - \cM_{1'2'}^2)^2 } g'_{t \, 2 \, q \bar q}(121'2')
\
b_1^\dagger d_2^\dagger d_{2'} b_{1'}    
\es
\left[ [\cH_f, \tilde \cH_{t\,1}], \cH_{t\,1} \right]_{q \bar q} \
.
\eeq
The first-order operator $\cH_{t\,1}$ is a sum of the two 
terms, $\cH_{t \, 1} = \cH_{t \, 1A} + \cH_{t \, 1B}$.
The terms  $\cH_{t \, 1A}$ and $\cH_{t \, 1B}$ describe 
the coupling of fermions to bosons of type $A$ and $B$, 
respectively. Their forms are identical to the ones given 
in Eqs.~(\ref{HpsiApsit}) and (\ref{HpsiBpsit}), except that 
the operators $q_t$ are replaced by $q$. The operator 
$\tilde \cH_{t \, 1}$ differs from $\cH_{t \ 1}$ by multiplication 
of its terms by the square of total $p^+$ of quanta annihilated 
or, equivalently, created by a term. Since the boson creation 
and annihilation operators must be contracted with each 
other on the right-hand side of Eq.~(\ref{cGprime}), one has
\beq
\label{cGprime1}
\sum_{121'2'}\int[121'2'] \,
\tilde \delta_{12.1'2'} \
e^{-t (\cM_{12}^2 - \cM_{1'2'}^2)^2 } g'_{t \, 2 \, q \bar q}(121'2')
\
b_1^\dagger d_2^\dagger d_{2'} b_{1'}    
\es
\left[ [H_f, \tilde \cH_{t\,1A}], \cH_{t\,1A} \right]_{q \bar q}+
\left[ [H_f, \tilde \cH_{t\,1B}], \cH_{t\,1B} \right]_{q \bar q}
.
\nn
\eeq
The result of integrating this 
equation, in compliance with the general RGPEP rules~\cite{RGPEP}, 
reads
\beq
\label{gintegrated}
g_{t \, 2 \, q \bar q}(121'2') - g_{0 \, 2 \, q \bar q}(121'2')
\es
\left[ e^{t(\cM_{12}^2 - \cM_{1'2'}^2)^2 } \ 
f_{t+t_r \, 12.x} \ f_{t+t_r \, x.1'2'} - f_{t_r \, 12.x} \ f_{t_r \, x.1'2'} \right]
\nt
{           p_{12.x}^+ a_{12.x} + p_{x.1'2'}^+ b_{x.1'2'}
\over
           (\cM_{12}^2 - \cM_{1'2'}^2)^2 
           - a_{12.x}^2
           - b_{x.1'2'}^2 }
\nt
\left[ \cH_{0\,1A \, 12.x} \cH_{0\,1A \, x.1'2'} + \cH_{0\,1B \, 12.x} \cH_{0\,1B \, x.1'2'} \right]_{12.1'2'} \ .
\eeq
The subscript $x = q b \bar q$ denotes the intermediate quanta. 
The momentum $p^+_{a.b}$ stands for the total $p^+$ of 
quanta that participate in the interaction caused by one operator 
$\cH_{0\, 1}$. The verex form factors are
\beq
f_{t \, 12.x} \es e^{- t a_{12.x}^2} \ , \quad a_{12.x} \rs p_{12.x}^+(P_{12}^- - P_x^-) \ , \\
f_{t \, x.1'2'} \es e^{- t b_{x.1'2'}^2}  \ , \quad b_{x.1'2'} \rs p_{x.1'2'}^+(P_{1'2'}^- - P_x^-) \ .
\eeq
Description of the resulting Hamiltonian coefficients
$h_{t \, 2 \, q \bar q}(121'2')$ in Eq.~(\ref{h2-121'2'}), 
will be provided in the next section after we introduce 
the additional terms that also contribute to the $q\bar q$ 
bound-state dynamics. 

\subsection{ Bound-state dynamics }
\label{boundstatedynamics}

The Hamiltonian $H_t$ determines the structure of
bound states (BS) through the eigenvalue equation
\beq
\label{HBS}
H_t \ |BS\rangle \es {P_{BS}^{\perp 2} + M_{BS}^2 \over P^+_{BS} } \ |BS\rangle \ .
\eeq
The kinematic total bound-state momentum components
$P_{BS}^\perp$ and $P_{BS}^+$ can be eliminated since
the FF of Hamiltonian dynamics and the RGPEP both 
explicitly preserve the seven Poincar\'e symmetries
that include the Lorentz boosts. Therefore, $P_{BS}^+$
and $P_{BS}^\perp$ are arbitrary and only the eigenvalue
$M_{BS}^2$ needs to be found. The relative motion of 
constituents is described in terms of the wave functions 
that do not depend on $P_{BS}^+$ and $P_{BS}^\perp$.
Therefore, the same wave functions appear in the bound-state
spectroscopy and in the corresponding parton picture
in the infinite momentum frame~\cite{FeynmanPM}.
However, the wave functions depend on the constituent
size $s = t^{1/4}$. Therefore, the parameter $t$ plays
the role of scale of constituents one uses to describe the
bound state. An external probe may couple differently to 
constituents of different size, as is the case in the 
electro-weak form factors, deep inelastic scattering
or virtual Compton scattering.

In the case of bound states of a fermion and an anti-fermion,
the wave functions appear in the expansion 
\beq
\label{Bexp}
|BS\rangle \es
\sum_{q \bar q} \psi_{t \, q \bar q} |q_t \bar q_t \rangle
+
\sum_{q b \bar q} \psi_{t \, q b \bar q} |q_t b_t \bar q_t \rangle
+
. . . \ ,
\eeq
where the sum extends to infinite numbers of effective fermion, 
anti-fermion and boson quanta. In a local gauge theory, the 
integrals over constituent momenta extend to infinity and the 
expansion is hardly convergent~\cite{Dyson}. In the effective 
theory with constituents of size $s$, approached here using the 
RGPEP, the convergence is conceivable because the ultraviolet range 
of interactions is limited by the vertex form factors $f_{c.a}$, 
see Eq.~(\ref{ftc.a}). The infrared divergences due to massless 
gauge bosons~\cite{BlochNordsieck,Nordsieck}, are tamed 
by the introduction of mass $\kappa$ and an additional polarization
state. The mass $\kappa$ also appears in the form factors 
$f_{t \, c.a}$, which thus tame small-$x$ singularities in 
dynamical considerations that concern partons~\cite{FeynmanPM}. 

When the coupling constant is very small, one may attempt to 
solve Eq.~(\ref{HBS}) by assuming that the smallest eigenvalue
$M_{BS}^2$ corresponds to the state dominated by its fermion-anti-fermion
component in Eq.~(\ref{Bexp}). The component with one boson is
of order $g$ and the remaining components are of order $g^2$
or smaller. For example, such approach can be adopted in QED,
where $g$ is the electron electric charge. Expansion in powers of 
$g$ allows one to derive an effective Hamiltonian matrix that acts 
solely on the wave functions $\psi_{t \, q \bar q}$ in the space of 
fermion-anti-fermion components $|q_t \bar q_t \rangle$. We use 
the second-order formula~\cite{WilsonR}
\beq
\label{Heffqbarq}
\langle 1_t2_t| H_{t \, {\rm eff} \, 2 \, q \bar q} | 1'_t 2'_t \rangle
\es
\langle 1_t 2_t| H_{t \, 2}|1'_t 2'_t \rangle
+
\2 \sum_{ x \neq q \bar q} 
\left( {1 \over P_{12}^- - P_x^-} + {1 \over P_{1'2'}^- - P_x^-} \right)
\langle 1_t 2_t| H_{t \, 1} | x_t\rangle \langle x_t| H_{t \, 1} |1'_t 2'_t \rangle \ .
\eeq
On the right-hand side, there are six kinds of terms 
due to the operator $H_{t \, 2}$ and similar six kinds 
of terms due to the term bilinear in $H_{t\,1}$. The latter
terms are the effective self-interaction of fermions, self-interaction 
of anti-fermions, exchange of bosons of types $A$ 
and $B$ between fermions, and annihilation of 
fermion-anti-fermion pairs into the two types of bosons 
with subsequent creation of a fermion pair. Note that 
the Hamiltonian $H_{t\,2}$ whose matrix elements appear 
as the first term on the right-hand side of Eq.~(\ref{Heffqbarq}), 
results from a solution of differential Eq.~(\ref{h2prime1}) 
for a Hamiltonian operator that acts in the entire Fock 
space. In contrast, the term bilinear in Hamiltonians 
$H_{t \, 1}$ only describes the interactions of effective 
particles in the fermion-anti-fermion component of the 
bound-state eigenvalue problem for small values
of $M_{BS}$. In other words, the matrix element 
$\langle 1_t2_t| H_{t \, {\rm eff} \, 2 \, q \bar q} | 1'_t 2'_t \rangle$
corresponds to an operator $H_{t \, {\rm eff} \, 2 \, q \bar q}$
that acts solely in the effective fermion-anti-fermion sector 
of Fock space, built from quanta of size $s$ for description  
of bound-states dominated by that component.

\subsection{ Eigenvalue problem for bound state wave functions }
\label{bseigenvalue}

The bound-state eigenvalue problem of Eq.~(\ref{Bexp}),
reduced to the dominant fermion-anti-fermion component reads
\beq
\label{BSeigenvalueEquation}
(p_1 + p_2)^2 \ \psi_{t \, 12}
+
g^2 P_{BS}^+ \sum_{1'2'}\int[1'2'] \
\langle 1_t 2_t| H_{t \, {\rm eff} \, 2 \, q \bar q} | 1'_t 2'_t \rangle \ \psi_{t \, 1'2'}
\es
M_{BS}^2 \ \psi_{t \, 12} \ .
\eeq
The mass corrections $\delta m^2(t)$ are canceled by 
the effective fermion self-interactions due to the term bilinear 
in $H_{t \, 1}$ in Eq.~(\ref{Heffqbarq}). The invariant mass
squared of two constituent fermions, $\cM_{12}^2 = (p_1+p_2)^2$,
is calculated using on-mass-shell values of $p_1^-$ and $p_2^-$
with fermion mass eigenvalue $m$. The whole interaction left 
consists of the exchange and annihilation terms. They involve 
sums over polarizations of bosons of type $A$ and $B$. The 
sums result in tensors $d_{A \, \mu \nu}$ and $d_{B \, \mu \nu}$ 
that are contracted with the fermion currents $j^\mu_q$ 
and $j^\nu_{\bar q}$. The transverse boson tensor 
$d_{A \, \mu \nu}$ includes the metric term $-g_{\mu \nu}$
and an additional tensor that involves the boson momentum. 
Using conservation of kinematic momentum components and
properties of spinors in the fermion currents, one can reduce 
the additional tensor to $\eta_\mu \eta_\nu$ times a coefficient, 
where the four-vector $\eta$ has only minus component different 
from zero, and equal two. The tensor $d_{B \, \mu \nu}$ is  
proportional to $\eta_\mu \eta_\nu$. The second-order 
interaction matrix in $\langle 1_t 2_t| H_{t \, {\rm eff} \, 2 \, q \bar q} 
| 1'_t 2'_t\rangle$ thus takes the form  
\beq
\label{effhfbarfadding}
\langle 12| H_{t \, {\rm eff} \, 2 \, q \bar q} | 1'2'\rangle
\es
\tilde \delta_{12.1'2'}  
\ h_{t \, {\rm eff} \, 2 f \bar f}(121'2') \ ,
\eeq
where
\beq
\label{4lines}
h_{t \, {\rm eff} \, 2 f \bar f}(121'2') \es L_1 + L_2 + L_3 + L_4 \ , \\
\label{4lines EXg}
L_1 \es EX_g \left[ h_{g_{\mu \nu} \, {\rm exch}} + h_{g_{\mu \nu} \,
{\rm boson ~ exch} } \right] \ , \\
\label{4lines EX+}
L_2 \es
EX_+ \left[ h_{\gamma^+ \, {\rm exch}} + h_{\gamma^+ \, {\rm
boson ~ exch}} \right] \ , \\
\label{4lines ANg}
L_3 \es
AN_g \left[ h_{g_{\mu \nu} \, {\rm annih}} + h_{g_{\mu \nu} \,
{\rm boson ~ annih}} \right] \ , \\
\label{4lines AN+}
L_4 \es
AN_+ \left[ h_{\gamma^+ \, {\rm annih}} + h_{\gamma^+ \, {\rm
boson ~ annih}} \right] \ ,
\eeq
and the spinor factors are  
\beq
\label{EXg}
EX_g \es - \ \bar u_1 \gamma^\mu u_{1'} \ \bar v_{2'} \gamma_\mu
v_2 / (2m)^2 \ , \\
\label{EX+}
EX_+ \es \bar u_1 \gamma^+ u_{1'} \ \bar v_{2'} \gamma^+ v_2 /
(p_1^+ + p_2^+)^2 \ , \\
\label{ANg}
AN_g \es - \ \bar u_1 \gamma^\mu v_2 \ \bar v_{2'} \gamma_\mu
u_1 / (2m)^2 \ , \\
\label{AN+}
AN_+ \es \bar u_1 \gamma^+ v_2 \ \bar v_{2'} \gamma^+ u_{1'} /
(p_1^+ + p_2^+)^2 \ .
\eeq
The terms with subscripts ``exch'' or ``annih'' come from 
the operator $H_{t \, 2}$, and terms with subscripts
``boson exch'' or ``boson annih'' from the term bilinear
in $H_{t\, 1}$ in Eq.~(\ref{Heffqbarq}). Our results for 
the four terms in Eq.~(\ref{4lines}), denoted by $L_1$, 
$L_2$, $L_3$, $L_4$ and called ``lines'', are listed below.
The coupling constant square $g^2$ does not appear in
them since it is factored out in Eq.~(\ref{BSeigenvalueEquation}).
Each of the lines consists of a dimensionless spinor factor 
and a dimensionless function of fermions' momenta in a 
square bracket. The latter functions will be called {\it relativistic 
potentials} for two reasons. One reason is that the functions 
are invariant with respect to the seven kinematic Poincar\'e 
transformations of FF dynamics that include boosts. The 
other reason is that the corresponding Hamiltonian interaction 
terms do not change the number of effective particles. 
Below, the relativistic potentials in lines $L_1$ to $L_4$ 
are for brevity called just potentials and denoted by $V_1$ 
to $V_4$, respectively. Note the negative signs in front 
of spin factors in lines $L_1$ and $L_3$. Thus, for small 
relative momenta of fermions, a positive potential $V_1$ 
implies attraction and positive potentials $V_2$, $V_3$ 
and $V_4$ imply repulsion. All these potentials are 
dimensionless functions of kinematical momenta of 
four fermions, their mass, the mass of gauge bosons
and the scale parameter $s$.

There are no counter terms included in the lines listed below, 
because none is needed. Matrix elements of the interaction 
terms between wave packets of fermions~\cite{Wilsonetal} 
do not depend on the regularization parameter $t_r$ in the 
limit $t_r \to 0$. The interaction ultraviolet behavior is 
limited by the RGPEP form factors with finite parameter $t$.
Fermions have masses and do not produce any infrared
singularities. The infrared singularities due to the bosons  
are regulated by the mass $\kappa$ and small-$x$ 
singularities for finite effective-particle size $s$ are 
removed by the lower bound on the boson $x$ on the order
of $s^2 \kappa^2$. In addition, the logarithmic dependence 
on that bound cancels out in the sense of principal value 
in the integrals with wave packets. More details are reported 
in Secs.~\ref{relativisticpotentials} and \ref{plotsofrelativisticpotentials}.

\subsection{ Relativistic potentials }
\label{relativisticpotentials}

In the list of interaction terms in lines $L_1$ to $L_ 4$, we use 
the familiar parton-model parameterization of constituents'
momenta, commonly used in the literature that employs FF 
dynamics,
\beq
\label{p+BS}
p_{1,2}^+ \es x_{1,2} P_{BS}^+ \ , \\
\label{pprime+BS}
p_{1',2'}^+ \es x_{1',2'} P_{BS}^+ \ , \\
\label{pperpBS}
p_{1,2}^\perp \es x_{1,2} P_{BS}^\perp \pm k^\perp \ , \\
\label{pprimeperpBS}
p_{1',2'}^\perp \es x_{1',2'} P_{BS}^\perp \pm k'^\perp \ .
\eeq
We also use the abbreviation $z = x_{1'} - x_1$ and
introduce two four-momentum transfers for fermions, 
\beq
q_1 \es p_{1'} - p_1 \ , \\
q_2 \es p_2 - p_{2'} \ .
\eeq
These differ only in their minus components, evaluated 
using the on-mass-shell fermion four-momenta. The 
relativistic potentials are expressed in terms of quantities
analogous to a denominator $d = \kappa^2 - p^2$ in 
the Feynman propagator for bosons,
\beq
d_1 \es \kappa^2 - q_1^2 \ , \\
d_2 \es \kappa^2 - q_2^2  \ .
\eeq
Four invariant-mass quantities are introduced for brevity, 
\beq
a \es \cM_{1 2}^2  - m^2 \ , \\
a' \es \cM_{1'2'}^2  - m^2 \ , \\
b \es \cM_{1 2}^2  - \kappa^2 \ , \\
b' \es \cM_{1'2'}^2 - \kappa^2 \ .
\eeq
All potentials are listed below ignoring the regularization
parameter $t_r$. The RGPEP form factors with finite $t$
mute the presence of $t_r$ as negligible in comparison 
with $t$ in the sum $t_r + t$.

In the line $L_1$ of Eq.~(\ref{4lines EXg}), written in the form 
\beq
L_1 \es EX_g \ V_1 \ ,
\eeq
the relativistic potential reads
\beq
\label{potentialV1}
V_1(121'2') \es h_{g_{\mu \nu} \, {\rm exch}} + h_{g_{\mu \nu} \, {\rm boson ~ exch} } 
\rs
\theta(z) \ T_1  +  \theta(-z) \ T_2 \ , 
\eeq
where
\beq
T_1 \es T_{1f} \ e^{-t \, (a-a')^2 } + T_{1ff} \ e^{-t \, ( d_1^2 x_{1'}^2 + d_2^2 x_2   ^2 ) / z^2 } \ , \\
T_2 \es T_{2f} \ e^{-t \, (a-a')^2} + T_{2ff} \ e^{-t \, ( d_1^2 x_1    ^2 + d_2^2 x_{2'}^2 ) / z^2 } \ , 
\eeq
and
\beq
\label{line1}
T_{1f} 
\es 
{ 4m^2 ( d_1 x_{1'}^2 + d_2 x_2    ^2 ) \over d_1^2 x_{1'}^2 + d_2^2 x_2^2    - (d_2 - d_1)^2 } 
\ , \\
T_{2f}  
\es
{ 4m^2 ( d_1 x_1    ^2 + d_2 x_{2'}^2 ) \over  d_1^2 x_1^2    + d_2^2 x_{2'}^2 - (d_2 - d_1)^2 } 
\ , \\
T_{1ff} 
\es
2 m^2/ d_2 + 2m^2/d_1 
\ - \ 
T_{1f} 
\ , \\
T_{2ff} 
\es
2m^2/d_1 + 2m^2/d_2 
\ - \ 
T_{2f} 
 \ .
\eeq
Note that $a-a' = (d_1 - d_2)/z$. For small momentum transfers, 
line $L_1$ provides a Yukawa potential due to the exchange of 
vector bosons of mass $\kappa$ between fermions, including the 
familiar spin factors. However, off-shell, {\it i.e.}, when the invariant 
mass of fermions before the interaction differs from their invariant 
mass after the interaction, $a \neq a'$, the potential's behavior is 
quite different from the commonly known one in the non-relativistic 
Schroedinger equation. Further discussion is provided in 
Secs.~\ref{spectroscopyandPMpictures} and \ref{plots}. 

The relativistic potential in line $L_2$ of Eq.~(\ref{4lines EX+}), 
written in the form 
\beq
\label{L2EX+V2}
L_2 \es EX_+ \ V_2 \ ,
\eeq
is
\beq
\label{potentialV2}
V_2(121'2') \es
h_{\gamma^+ \, {\rm exch}} + h_{\gamma^+ \, {\rm boson ~ exch}}
\rs
[ \theta(z) \ S_1 + \theta(-z) \ S_2 ] (d_1 - d_2)/z^2
\ , 
\eeq
where
\beq
S_1 \es S_{1f} \ e^{-t \, (a-a')^2 } + S_{1ff} 
\ e^{-t \, ( d_1^2 x_{1'}^2 + d_2^2 x_2   ^2 ) / z^2 } \ , \\
S_2 \es S_{2f} \ e^{-t \, (a-a')^2 } + S_{2ff} 
\ e^{-t \, ( d_1^2 x_1    ^2 + d_2^2 x_{2'}^2 ) / z^2 } \ , 
\eeq
and
\beq
S_{1f}
\es \2
{  -  d_1 x_{1'}^2  + d_2 x_2^2    + 2(d_1 - d_2) \over  
d_1^2 x_{1'}^2 + d_2^2 x_2^2 - (d_1 - d_2)^2} \ , \\
S_{2f}
\es \2
{   - d_1 x_1^2      + d_2 x_{2'}^2  + 2(d_1 - d_2) \over  
d_1^2 x_1^2 + d_2^2 x_{2'}^2 - (d_1 - d_2)^2} \ , \\
\label{T1ffpom}
S_{1ff}
\es \4
 {    d_1^2 x_{1'}^2 - d_2^2 x_2^2  - d_1^2  + d_2^2  \over 
d_1^2 x_{1'}^2 + d_2^2 x_2^2 - (d_1 - d_2)^2 } 
\ (1 /d_2+ 1/d_1)  \ , \\
S_{2ff}
\es \4
 { d_1^2 x_1^2   -  d_2^2 x_{2'}^2  - d_1^2  + d_2^2  \over 
d_1^2 x_1^2 + d_2^2  x_{2'}^2 - (d_1 - d_2)^2 } 
 \ (1 /d_1 + 1/d_2) \ .
\eeq
Since $d_1 - d_2 = z (a-a')$, the potential $V_2$ is capable 
in the limit $z \to 0$ of behaving like $1/z$ and producing a 
singularity. However, the singularity is integrable with regular 
bound-state wave functions in the sense of principal value, 
{\it cf.}~\cite{principalvalue}. For small momentum transfers, 
one has $a \sim a'$ and the potential approaches a regular 
function near $z=0$. The entire potential $V_2$ vanishes on 
shell, {\it i.e.}, when the invariant masses of fermions before 
and after the interaction are the same, $a = a'$. Hence, $V_2$ 
does not contribute to the on-shell scattering of fermions in the 
Born approximation. Consequently, it does not have any classical 
counterpart and differs qualitatively in this respect from the 
Yukawa potential.

Our result for the annihilation channel relativistic potential 
in line $L_3$ in Eq.~(\ref{4lines ANg}), written as 
\beq
L_3 \es AN_g \ V_3 \ ,
\eeq
reads
\beq
\label{potentialV3}
V_3(121'2')
\es
h_{g_{\mu \nu} \, {\rm annih}} + h_{g_{\mu \nu} \, {\rm boson ~ annih}} 
\rs
e^{ - t (b-b')^2 } \ { 4m^2(b+b') \over 2 \, b \, b'} \ .
\eeq
On shell, {\it i.e.}, when $b-b'=a-a'$ vanishes, our result for $V_3$ 
reduces to $4m^2/b$, which is fully covariant. From the line $L_4$ 
in Eq.~(\ref{4lines AN+}), written as 
\beq
\label{L4AN+V4}
L_4 \es
AN_+ \ V_4 \ ,
\eeq
we obtain the annihilation channel relativistic potential 
\beq
\label{potentialV4}
V_4(121'2') \es
h_{\gamma^+ \, {\rm annih}} + h_{\gamma^+ \, {\rm boson ~ annih}}
\rs
- \, e^{ -t(b-b')^2 }  {( b-b' )^2 \over 4 \, b \, b' } \ .
\eeq
Note the negative sign, which implies attraction. Potential $V_4$ 
vanishes on shell. It does not contribute to fermion-anti-fermion 
scattering matrix in the Born approximation.

\section{ Spectroscopy and the parton-model picture }
\label{spectroscopyandPMpictures}

This section provides a brief discussion that relates 
the computations described in previous sections to the
well-known physics of bound states in Abelian theory 
and their parton picture. The theory does not involve 
confinement. For the purpose of this discussion, we first 
need to clarify the relationship between the expansion 
in powers of $g$ used in the computations and the 
non-perturbative nature of the bound-state problem. 
The clarification is needed because the computed 
Hamiltonians only include terms of order 1, $g$ and 
$g^2$. As a consequence, the bound-state eigenvalue 
Eq.~(\ref{BSeigenvalueEquation}) does not contain 
interaction terms of higher order than second. 

The RGPEP usage of formal expansion in powers of $g$ 
does not mean that the bound states are described 
by perturbation theory. The actual situation is in this 
respect analogous to the situation in the original
non-relativistic Schroedinger equation in atomic 
physics~\cite{Schroedinger}. The Coulomb potential 
in that equation is just quadratic in the electric charge. 
Despite such low power of charge, the atomic bound 
states are successfully described using the Coulomb 
potential. They are not describable using perturbation 
theory. The critical step beyond perturbation theory is 
made by solving the eigenvalue problem for the 
Hamiltonian. Similarly, the second-order RGPEP leads 
to Eq.~(\ref{BSeigenvalueEquation}) that is capable 
of describing bound-state wave functions as 
non-perturbative objects.

We wish to stress at this point that the RGPEP computation 
can also be carried out in expansion to higher orders than 
second. Results could suggest the structure of effective FF 
Hamiltonians needed to properly account for some non-perturbative 
effects of the theory. For examples of computing or guessing 
such terms, see~\cite{Wilsonetal,KSerafinB,KSerafinPhD} 
and references therein. It is also worth stressing that the 
Hamiltonians computed using the RGPEP are obtained without 
putting any restriction on the motion of field quanta and without 
making any non-relativistic approximation concerning their 
motion. This is relevant to our discussion because for self-evident 
reasons the connection between spectroscopy and parton picture 
for bound states cannot be rigorously formulated in a 
non-relativistic theory.

Suppose that a wave function $\psi_{t \, 12}$ is 
a solution not of Eq.~(\ref{BSeigenvalueEquation})
but of the analogous eigenvalue equation that is 
derived by first solving the RGPEP Eq.~(\ref{RGPEP}) 
for $\cH_t$ exactly and subsequently reducing 
the eigenvalue problem for $H_t$ to the bound-state 
dominant effective Fock-space component eigenvalue 
equation also exactly, instead of using expansions in 
powers of $g$ that we used to derive 
Eq.~(\ref{BSeigenvalueEquation}). The exact wave 
functions $\psi_{t \, 12}$ would describe the bound 
states in terms of the effective constituents of scale 
$s = t^{1/4}$. Using the analogy with the Schroedinger 
equation, one would then expect that the spectroscopy 
of bound states could be developed in terms of such 
constituents and their wave functions. The wave 
functions could be used for calculating bound-state
observables.

As an example of a bound-state observable, consider 
scattering of electrons off a bound state. It is 
characterized by the momentum transfer $Q$ and 
possibly other parameters, such as the Bjorken
$x$ in deep inelastic scattering (DIS). The cross section 
in DIS will involve the bound-state's structure functions. 
The cross section in the elastic scattering will involve 
the bound-state's form factors, {\it etc.} Once the wave 
functions are known, the observables can be studied 
using familiar FF formulas~\cite{FFreview1,FFreview2,
FFreview3,FFreview4,FFreview5,FFreview6}. However,
the Hamiltonian interaction terms one could so use  
apply for the effective constituents of size $s$, instead 
of the abstract, point-like quanta of canonical theory. 

Calculation of the bound-state observables will produce 
results that depend on the scale $Q$ and other parameters, 
such as $x$. The dependence will result from the kinematics 
and dynamics of the constituents of size $s$. As in other 
approaches, {\it e.g.}, see~\cite{PMC1,PMC2}, one expects 
that the calculation will take the simplest form when the constituent 
size $s$ will be optimized for the purpose. For example,
setting $s = 1/Q$ or $s =\sqrt{x} /Q$ makes the 
corresponding logarithms of the products $sQ$ or $s^2 
Q^2/x$ vanish. In other words, although the size of constituents 
does not influence the values of observables, since it 
plays the role of a renormalization group parameter in
the full theory that is not limited to any perturbative expansion, 
the choice of $s$ does influence the complexity of calculation.

When the Hamiltonian $H_t$ and associated effective few-body 
interactions, are derived using the RGPEP  in a perturbative 
expansion, which is the case in Eq.~(\ref{BSeigenvalueEquation}), 
there will be residual dependence of calculated bound-state 
observables on the size of effective quanta $s$. This 
dependence should be reduced by using the running coupling 
constant $g_t$ as the expansion parameter for description 
of phenomena of scale $s$.

Connection between the bound-states' spectroscopy developed 
in terms of the wave functions such as $\psi_{t \, 12}$ in 
Eq.~(\ref{BSeigenvalueEquation}), and the bound-states' 
features, such as parton distributions, is based on the following 
observations. When the coupling constant $g$ is very small, the 
dominant interaction term in Eq.~(\ref{BSeigenvalueEquation}) 
is the Yukawa potential that for an extremely small boson mass is 
practically equivalent to the Coulomb potential. Namely, when
one denotes by $\vec k$ the relative momentum of fermions
1 and 2 and by $\vec k\,'$ the relative momentum of fermions
1' and 2', using the relative three-momentum variables defined 
in App.~\ref{notation}, then the dominant interaction term in 
Eq.~(\ref{effhfbarfadding}), through Eq.~(\ref{potentialV1}), 
takes the form
\beq
\label{FFCoulomb}
h_t  (121'2') 
\es
-\, e^{ - 16 t (\vec k\,^2 - \vec k\,'\,^2)^2 }  
{ 4m^2 \over (\vec k - \vec k\,' )^2 + \kappa^2 } \ .
\eeq
The boson mass can be extremely small. For the values of $t$ 
that correspond to the size $s$ much smaller than the Bohr 
radius of the system, the RGPEP form factor in front can be 
ignored and we obtain a picture that closely resembles the 
non-relativistic Schroedinger equation for positronium. In such 
a system, the concept of spectroscopy is well understood.

The associated parton picture is obtained on the basis 
of observation that the relative momentum variables 
in the FF of Hamiltonian dynamics are invariant 
with respect to boosts. The bound-state wave function 
$\psi_{t \, 12}= \psi_t(\vec k \,)$ as a function of 
variables $x$ and $k^\perp$, see App.~\ref{notation},
\beq
x \es (1 + k^z/E_k)/2 \ , \\
E_k \es \sqrt{ m^2 + \vec k\,^2} \ ,
\eeq
is the same in the bound-state rest frame as in the 
infinite momentum frame (IMF). Therefore, the 
wave function $\psi_{t \, 12} = \phi_t(x,k^\perp)$ 
provides the probability distribution $f(x)$ of 
constituents as partons in the IMF,
\beq
\label{fx}
f(x) & \sim & \int d^2k^\perp \  \phi_t^2(x,k^\perp ) \ .
\eeq
In the integrals over relative motion of constituents or
partons, one has to also keep track of minimal relativity
factors indicated in Eq.~(\ref{minimalrelativity}) in 
App.~\ref{notation}. The main point is, however, that the 
size $s$ of the constituents plays the role of scale parameter. 
Our computations in the previous sections need to be 
improved by including variation of the coupling constant 
$g_t$ with $t$, {\it cf.}~\cite{Gomez}. Moreover, according 
to Eq.~(\ref{Uqt}), the operators for quanta corresponding 
to different scales are related by a unitary operator 
$W_{t_1 t_2} = \cU_{t_1} \cU_{t_2}^\dagger$,
\beq
\label{Uqt1}
q_{t_1} \es W_{t_1 t_2}  q_{t_2} W_{t_1 t_2} ^\dagger \ .
\eeq 
The parton distributions obtained from the wave functions
such as $\phi_t(x,k^\perp )$ will vary with $t$ due to the
effects of fermions emitting bosons, bosons splitting into 
fermion pairs and the corresponding reverse processes. These effects are hidden 
in the transformation $W_{t_1 t_2}$, which is computable 
order-by-order using the RGPEP~\cite{GlazekCondensatesAPP,TrawinskiPhD}. 
The transformation $W_{t_1 t_2}$ relates the field quanta 
of size $s_1$ that most efficiently describe the binding-mechanism, 
to the field-quanta  of size $s_2$ that the external 
probe is most sensitive to. 

The bound-state eigenvalue problem of Eq.~(\ref{HBS}) 
for the Hamiltonian $H_t$, will also lead to the intrinsic 
Fock-space components of the eigenstates written in
terms of constituents or partons of size $s$. These intrinsic 
components are not described just by the RGPEP evolution 
operator $W_{t_1 t_2}$, but by the non-perturbative 
solutions to the eigenvalue problem. In handling these
components using perturbation theory, one needs to be
careful in order to avoid double counting.

In summary, the RGPEP opens a way for seeking a connection 
between the spectroscopy of bound-states with their 
parton-distribution picture. Most succinctly, one could say 
that the present formulation of Abelian gauge theory, with 
the gauge boson mass introduced as a regulator of infrared
and small-$x$ divergences, provides a partial hint on seeking 
a ``satisfactory method of truncating the theory'' to identify 
the binding mechanism of constituent quarks and 
partons~\cite{Melosh,GellMann}.

\section{ Plots of masses and potentials }
\label{plots}

This section provides plots that illustrate the Hamiltonian 
mass correction and potential interaction terms that are 
computed in Secs.~\ref{fermionselfinteractions}, 
\ref{bosonselfinteractions} and \ref{relativisticpotentials}.
Plots of corrections to masses squared may appear
superfluous to some extent because the self-interactions 
of effective quanta cancel them precisely. However, the
plots show the orders of magnitude of the terms that 
cancel out. Their magnitude raises questions about 
formal applicability of perturbation theory for realistic 
values of the coupling constant, which we shall comment 
on. Regarding the interactions between fermions, plots 
that illustrate the effective one-boson-exchange interaction 
and the interaction in the annihilation channel, show in
what way and how much the quantum off-shell dynamics 
of effective quanta differ from the non-relativistic Schroedinger 
equation with the Coulomb or Yukawa potential. 

\subsection{ Mass corrections }
\label{masscorrections}

As a result of quantitative control on ultraviolet and infrared
singularities through the RGPEP and gauge-boson mass 
parameter $\kappa$, one can plot the behavior of mass 
corrections in the Hamiltonian $H_t$. Note that $\kappa$ 
is {\it a priori} arbitrary and can be made extremely small 
simultaneously with lifting the regularization. The latter is
done by making the regularization parameter $t_r$ negligible
in comparison with the finite RGPEP parameter $t$. After 
carrying out integration over transverse momentum in the 
mass-correction formulas given in Eqs.~(\ref{deltam2t}), 
(\ref{kappaApom}) and (\ref{kappaBpom}), we obtain 
\beq
\label{deltamplot}
g^2 \delta m^2(t) 
\es
{\alpha_g \over 4 \sqrt{2\pi} } 
\ 
{I_{FE}(t) \over \sqrt{t+t_r}} 
-
{\alpha_g \over 4 \pi}
\left( 2 m^2 + \kappa^2 \right) 
\ 
I_{FG}(t) 
\ , \\
\label{kappaAplot}
g^2 \delta \kappa_A^2(t) 
\es
{\alpha_g \over 4 \sqrt{2\pi} } 
\ 
{I_{AE}(t) \over \sqrt{t+t_r}} 
+
{\alpha_g \over 4\pi } 
\left( 2 m^2 + \kappa^2 \right) 
\ 
I_{AG}(t)
\ , \\
\label{kappaBplot}
g^2 \delta \kappa_B^2(t) 
\es
 {\alpha_g \over 4 \pi } \ \kappa^2 
\
I_{BG}(t)
\ ,
\eeq
where $\alpha_g=g^2/(4\pi)$ and the scale-dependent integrals are
\beq
\label{IFE}
I_{FE}(t) 
\es \int_0^1 dx \
{1 + (1-x)^2 \over x }   \ {\rm erfc}\left[ \sqrt{2(t+t_r)} \, \delta \cM_{fb}^2 \right] \ , \\
\label{IFG}
I_{FG}(t) 
\es \int_0^1 dx \
\Gamma \left[0, 2(t+t_r) \, \delta\cM_{fb}^4 \right] \ , \\
\label{IAE}
I_{AE}(t) 
\es \int_0^1 dx \
\left[  x^2 + (1-x)^2 \right]  \ {\rm erfc} \left[ \sqrt{2(t+t_r)}\, \delta \cM_{f \bar f}^2 \right] \ , \\
\label{IAG}
I_{AG}(t)
\es \int_0^1 dx \
\left[ 1 - { \kappa^2 x(1-x)  \over m^2 + \kappa^2/2 }\right]  
\ \Gamma \left[ 0,2(t+t_r) \, \delta \cM_{f \bar f}^4 \right] \ , \\
\label{IBG}
I_{BG}(t)
\es \int_0^1 dx \ 4 x(1-x) \
\Gamma \left[ 0,2(t+t_r) \, \delta \cM_{f \bar f}^4 \right] \ .
\eeq
Symbols erfc and $\Gamma$ denote the complementary error and 
incomplete gamma functions. They are referred to by the subscripts 
$FE$, $FG$, $AE$, $AG$ and $BG$ of the integrals, in correspondence
to fermion erfc, fermion gamma, boson A erfc, boson A gamma and 
boson B gamma. The degrees of off-shell departure of invariant-masses 
squared are 
\beq
\label{cMfbpom}
\delta \cM_{fb}^2 \es \kappa^2/x +  m^2/(1-x) - m^2 \ , \\
\label{cMffpom}
\delta \cM^2_{f \bar f} \es m^2/x + m^2/(1-x) - \kappa^2 \ .
\eeq
In the limit $t \to 0$, Eqs.~(\ref{deltamplot}), (\ref{kappaAplot}) 
and (\ref{kappaBplot}) provide the values of the mass-squared 
counter terms introduced in the initial, canonical Hamiltonian that 
is regulated using $t_r \to 0$. 
\begin{figure}[ht!]
          \caption{Five integrands of the integrals in Eqs.~(\ref{IFE}) to (\ref{IBG}) 
          that contribute to the fermion and boson effective masses, as indicated by 
          their subscripts, for four values of the effective particle size $s$. The 
          sequence shows how the integrands vary when the size $s$ is decreased. 
          The coupling constant $\alpha_g = 1/137$ and the gauge boson mass 
          is set equal to the fermion mass, $\kappa = m$. The corresponding values 
          of the mass corrections are given in Table~\ref{tab:table1}.}
          \label{fig:4plot}
          \includegraphics[width=.43\textwidth]{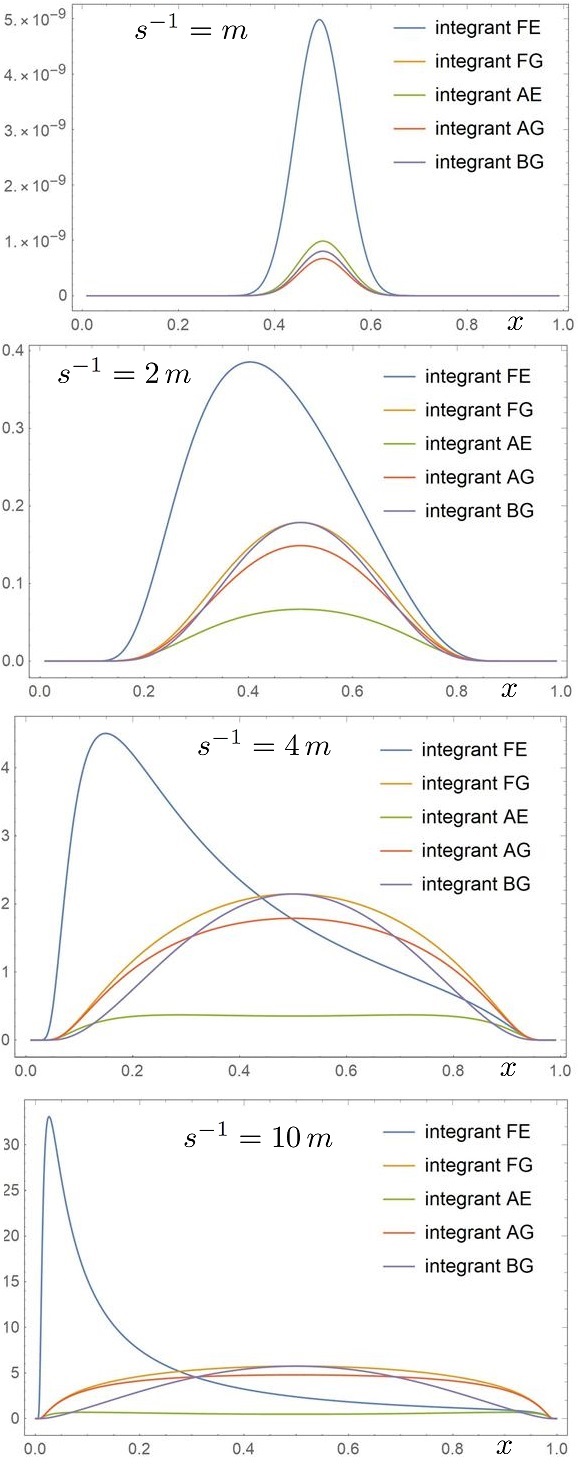}
          \end{figure}

\begin{figure}[ht!]
          \caption{The five integrands of the integrals in Eqs.~(\ref{IFE}) to 
          (\ref{IBG}) as indicated by their subscripts for two much smaller values 
          of the effective particle size $s$, necessarily in logarithmic scale, for 
          all other parameters without change. }
          \label{fig:2plotLog10}
          \includegraphics[width=.43\textwidth]{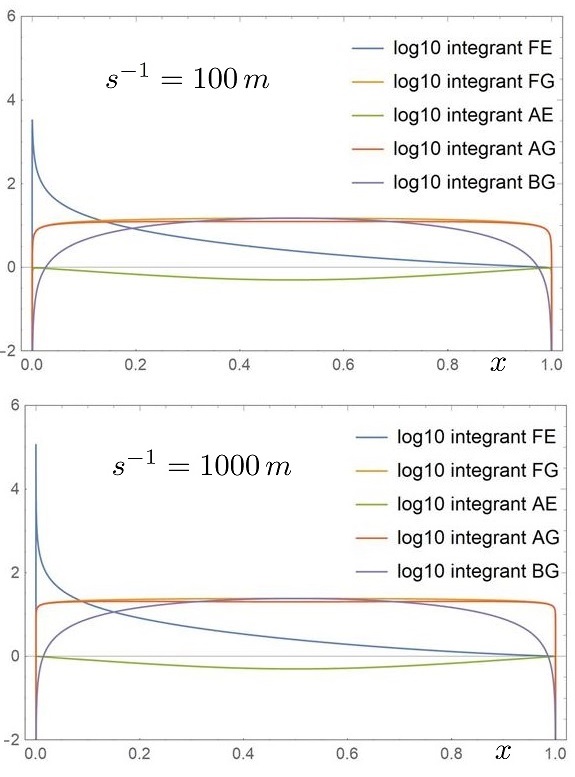}
          \end{figure}

For moderate values of $t$, the integrands of five integrals 
that contribute to the effective mass-squared corrections are 
plotted in Figs.~\ref{fig:4plot} and \ref{fig:2plotLog10}. The 
purpose of these figures is to show the origin of characteristic 
behavior of the mass-squared corrections as functions of the 
size of effective particles. For simplicity of the presentation 
and later discussion of what happens when the boson mass 
decreases, we set in these figures the boson mass $\kappa$ 
equal to the fermion mass $m$. The corresponding values 
of mass-squared corrections, all in ratio to $m^2$, are listed 
in Table~\ref{tab:table1}. We observe that the corrections 
are small for the size $s$ on the order of or greater than the 
Compton wavelength of fermions. The corrections grow quickly 
when $s$ decreases below the Compton wave length.

The fermion and transverse-boson (type $A$) mass-squared 
terms exhibit the dominant behavior $s^{-2}$. In contrast, the 
mass squared of longitudinal bosons (type $B$) is proportional 
to the physical value $\kappa^2$ and does not share with 
other quanta the rapid increase with $s^{-2}$. The 
fermion mass exhibits additional logarithmic increase with 
$s^{-2}$ due to the singular $x^{-1}$ behavior of the integral 
$I_{FE}$ for $x \to 0$, which is limited by the function erfc. 
The latter limits $x$ from below by a number order $s^2\kappa^2$, 
so the smaller $s$ the smaller allowed values of $x$ and the 
factor $1/x$ extends the support of fermion integrand toward 
$x=0$. In contrast, the boson mass integrands all behave symmetrically 
with respect to $x=1/2$. The difference between the fermion
and boson integrands originates in the first-order Hamiltonian 
interaction term that causes a fermion to emit a boson, which 
includes the factor $\sim 1/\sqrt{x}$ that is squared in $\delta m^2$. 
The boson mass-squared correction comes from the interaction 
that produces a fermion-anti-fermion pair, in which no such 
$x$-dependent, fast growing factor arises.
\begin{table}[ht!]
  \begin{center}
    \caption{ Values of mass corrections for equal boson and fermion masses, 
    $\kappa=m$, and six values of the size $s$ of effective fermion and boson 
    field quanta in units of the fermion Compton wavelength, according to 
    Eqs.~(\ref{deltamplot}), (\ref{kappaAplot}) and (\ref{kappaBplot}) 
    for $\alpha_g = 1/137$. The entries correspond to the integrands shown in 
    Figs.~\ref{fig:4plot} and \ref{fig:2plotLog10}. These corrections
    cancel out with the effective particle self-interactions. } 
    \label{tab:table1}
    \begin{tabular}{|c|c|c|c|c|c|c|}
\hline
$s \, m                               $ & $ 1                     $ & $ 0.5                  
                                         $ & $ 0.25                 $ & $ 0.1                   
                                         $ & $ 0.01                 $ & $ 0.001               $ \\      
\hline
$g^2\delta m^2/m^2         $ & $ 3.19 \ 10^{-13} $ & $ 3.03 \ 10^{-4}                   
                                         $ & $ 1.88 \ 10^{-2}   $ & $ 3.67 \ 10^{-1}   
                                         $ & $ 1.04 \ 10^{2}    $ & $ 1.71 \ 10^{4}  $ \\
\hline
$g^2\delta \kappa_A^2/m^2  $ & $ 2.73 \ 10^{-13} $ & $ 1.69 \ 10^{-4} 
                                             $ & $ 5.17 \ 10^{-3}  $ & $ 4.67 \ 10^{-2}   
                                             $ & $ 4.85                $ & $  4.85 \ 10^{2}   $ \\
\hline   
$g^2 \delta \kappa_B^2/m^2 $ & $ 5.64 \ 10^{-14} $ & $ 3.38 \ 10^{-5}                 
                                             $ & $ 6.21 \ 10^{-4}   $ & $ 1.96  \ 10^{-3}  
                                             $ & $ 5.52 \ 10^{-3}   $ & $ 9.09 \ 10^{-3}    $ \\
\hline         
    \end{tabular}
  \end{center}
\end{table}
In Fig.~\ref{fig:2plotLog10} the integrands are shown 
for values of $s$ hundred and thousand times smaller 
than the fermion Compton wavelength, approaching 
magnitudes comparable with the proton radius if the 
fermions have masses like electrons. The last two 
columns in Table~\ref{tab:table1} show how 
large the associated mass corrections become. The 
correction for fermions  grows much faster with $s^{-1}$ 
than the correction for bosons $A$ does. The mass 
correction for bosons $B$ is much smaller than for 
bosons $A$ and exhibits also much smaller rate of 
increase with $s^{-1}$. 

The large values of mass corrections for $\alpha_g = 1/137$
may rise readers' eyebrows. Indeed, such large values
suggest that the perturbative expansion is under suspicion
of inapplicability. However, the Hamiltonian $H_t$ leads
to the self-interactions of effective particles that cancel 
the large mass-squared corrections. One may hope that 
such precise cancellation among large terms survives in 
the non-perturbative solutions of the eigenvalue equations 
similar to Eq.~(\ref{BSeigenvalueEquation}). Indeed, once 
the large terms order $s^{-2}\log{s\kappa}$ and $s^{-2}$ 
are canceled by the effective particle self-interaction and
the remaining small parts are adjusted using eigenvalue 
equations for a single physical fermion and a single physical
boson, the bound-state equation for the fermion-anti-fermion
system is left with mass terms $m^2$ and $\kappa^2$ 
for all values of $t$. However, the warning that these results 
provide is that one needs a precise conceptual and quantitative 
control on the renormalized FF Hamiltonians, in order to describe 
binding of parton-like systems in gauge theories as well as one 
describes binding energies of constituents in spectroscopy of 
atomic systems.

\begin{figure}[ht!]
          \caption{Three integrands of the integrals in Eqs.~(\ref{IFE}) to (\ref{IAE}) 
          for the boson mass much smaller than the fermion mass, $\kappa = 10^{-25}\, m$
          for four values of the effective particle size $s$. The figure illustrates behavior 
          of the fermion integrand like $1/x$, where $x$ is the fraction of fermion momentum
          carried by the boson. Integrands in Eqs.~(\ref{IAG}) to (\ref{IBG}) are relatively
          so small that they cannot be shown on the figure. The coupling constant $\alpha_g 
          = 1/137$. The corresponding values of the mass corrections for fermions and 
          bosons are given in Table~\ref{tab:table1}.}
          \label{fig:4plotSmallKappa}
         \includegraphics[width=.38\textwidth]{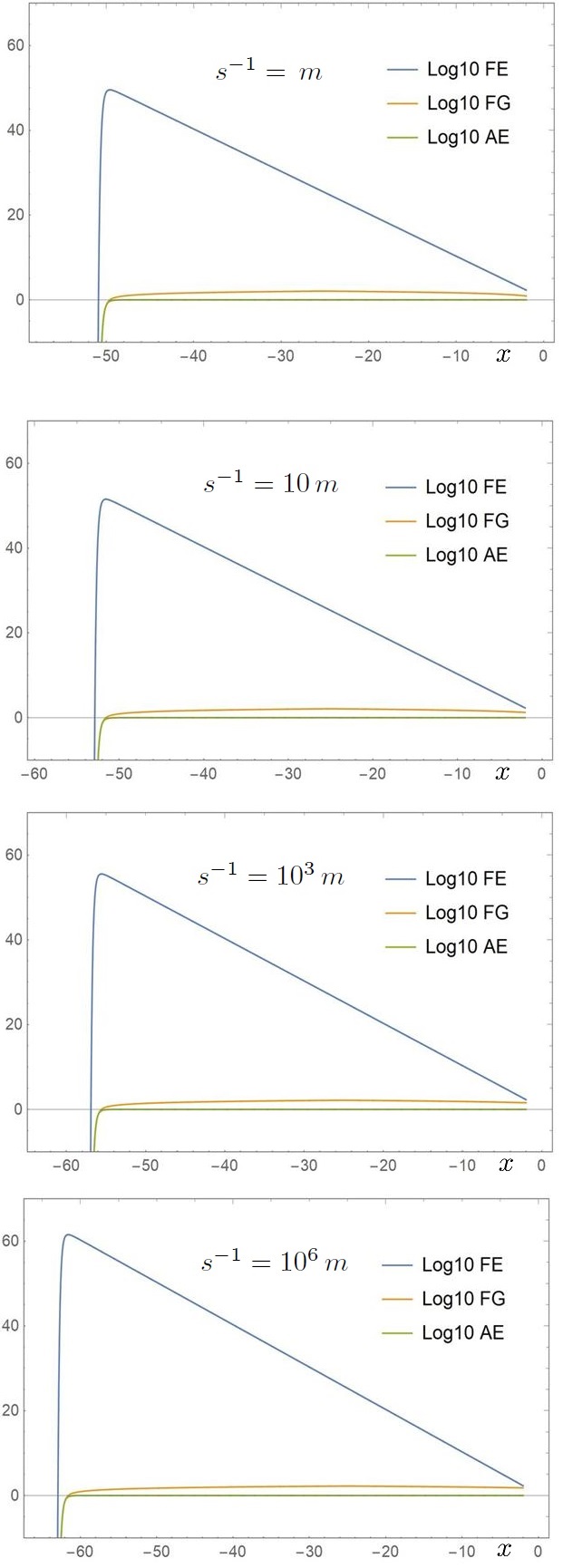}
          \end{figure}

In order to exhibit the actual magnitude of terms whose 
cancellation would have to be preserved, if one insisted 
on solving bound-state problems in canonical theory with 
some cutoff regularization that is meant to be lifted at the 
end of calculation, one can consider the gauge boson mass 
$\kappa = 10^{-18}$ eV. This is the currently accepted 
experimental upper bound on the photon mass~\cite{PDG}. 
In the computation, one can set $\kappa = 10^{-25} m$, 
imagining that $m$ could be the electron mass. On the 
basis of Figs.~\ref{fig:4plot} and \ref{fig:2plotLog10},
one can foresee the result. It is illustrated in Fig.~\ref{fig:4plotSmallKappa}
in terms of the plots of three integrands as functions of $x$.
Only three integrands are displayed because the 
remaining two are too small for showing them on 
the same  figure. Instead, Table~\ref{tab:table2} 
provides the resulting mass-squared corrections 
themselves, in ratio to the physical fermion mass.

The fermion mass correction is much larger than the 
boson mass corrections. One can see that it is logarithmically 
sensitive to the lower bound on $x$, which is effectively set 
by the RGPEP form factor to be around $\sqrt{2} s^2 \kappa^2$ 
divided by a number on the order of 100 or 1000. However, 
the dominant increase of the fermion mass correction  
is due to the factor $~s^{-2}$ that multiplies the logarithm.
The factor $s^{-2}$ is due to the integration over large 
transverse momenta of a boson with respect to a 
fermion in the intermediate state in fermion self-interaction. 

Boson masses behave differently. They do not exhibit
the logarithmic behavior in $s$ that fermions do because 
the intermediate states of the boson self-interaction only 
consist of fermion-anti-fermion pairs. The pair mass is 
$10^{25}$ times larger than the boson mass and the 
boson mass correction varies mostly due to the spinor 
factors that after integration over transverse momenta
render continuous and relatively slowly varying functions 
of $x$.

The intriguing feature of the boson mass corrections is 
that the types $A$ and $B$ are quite different, the latter
being very small in comparison to the former. This result 
can be confronted with the expectation that in the limit of 
$\kappa \to 0$ the third-polarization boson decouples from 
fermions because the coupling is proportional to 
$\kappa$~\cite{Soper,Yan3}. However, the actual 
coupling is of the form $\kappa/x$. Therefore, the small-$x$ 
behavior of the theory for $x$ order $s^2 \kappa^2$ or 
smaller includes contributions from the bosons of type $B$. 
Only after the cancellation of small-$x$ singularities for 
finite $s$, the limit $\kappa \to 0$ can be considered in 
quantum theory.

Concerning the magnitude of second-order mass corrections, 
we wish to state that in the case of constituent dynamics 
described by $H_t$ their values critically depend on the size 
of effective particles, see Tables~\ref{tab:table1} and \ref{tab:table2}.
When the size of effective fermions increases toward and above 
their Compton wavelengths, the magnitude of corrections rapidly 
decreases. For example, the entries in Table~\ref{tab:table2} for 
$sm=2$ would be from top to bottom $3.95\ 10^{-2}$, $1.40 \ 
10^{-5}$ and an incredibly small $7.01 \ 10^{-280}$. For $sm 
= 4$, we obtain, correspondingly, $9.63 \ 10^{-3}$, $1.05 
\ 10^{-6}$ and a number too small to quote. In Table~\ref{tab:table1}, 
increasing $sm$ to 2 results in mass corrections of order $10^{-131}$. 
If the RGPEP tendency for mass stabilization when $s$ crosses the 
fermion Compton wavelength survives in advanced computations, 
the models of bound states based on a few-body Schroedinger 
picture with potentials and practically fixed effective constituent 
masses could be adopted as a leading approximation. In the next 
section, we describe behavior of the second-order relativistic
potentials in a fermion-anti-fermion system.  
\begin{table}[ht!]
  \begin{center}
    \caption{ Values of mass corrections for the boson mass much smaller
                 than the fermion mass, $\kappa=10^{-25} \, m$, in agreement
                 with current experimental upper bound on the photon mass. 
                 Results for four values of the size $s$ of effective fermion and 
                 boson field quanta are shown in units of the fermion Compton 
                 wavelength, according to Eqs.~(\ref{deltamplot}), (\ref{kappaAplot}) 
                 and (\ref{kappaBplot}) for $\alpha = 1/137$. The entries correspond 
	                 to integrands shown in Fig.~\ref{fig:4plotSmallKappa}. These corrections
                  cancel out with the effective particle self-interactions. }
    \label{tab:table2}
    \begin{tabular}{|c|c|c|c|c|}
\hline
$s \, m                                  $ & $ 1                        $ & $ 10^{-1}                 
                                            $ & $ 10^{-3}              $ & $ 10^{-6}             $ \\      
\hline
$g^2\delta m^2/m^2            $ & $ 1.62 \ 10^{-1}    $ & $ 1.71  \  10^{+1}                  
                                            $ & $ 1.85 \ 10^{+5}   $ & $ 2.05 \ 10^{+11}  $ \\
\hline
$g^2\delta \kappa_A^2/m^2 $ & $ 1.39 \ 10^{-4}    $ & $ 4.93  \ 10^{-2} 
                                            $ & $ 4.85 \ 10^{+2}   $ & $ 4.85  \ 10^{+8}  $ \\
\hline   
$g^2 \delta \kappa_B^2/m^2 $ & $ 3.17 \ 10^{-70}  $ & $ 1.79  \ 10^{-53}                 
                                             $ & $ 8.92 \ 10^{-53}  $ & $ 1.96  \ 10^{-52} $ \\
\hline         
    \end{tabular}
  \end{center}
\end{table}

\subsection{ Plots of relativistic potentials }
\label{plotsofrelativisticpotentials}

The relativistic potentials for effective fermions of size $s$ 
are illustrated in this section by their action on wave functions 
of simple states. Consider a fermion-anti-fermion state described 
in terms of the parton-model variables. Let the fermions have 
equal momenta, so that they share their total momentum 
equally and their relative momentum is zero. To establish 
notation used for plotting potentials, this state of fermions is 
represented by
\beq
\label{state1'2'}
|1'2'\rangle \es b_{t \, 1'}^\dagger d_{t \, 2'}^\dagger |0\rangle \ ,
\eeq
where the individual momenta of fermions are $p_{1'} = p_{2'}=p'$
and their total momentum is $P_{1'2'} = 2p'$. We use labels 
with primes as in Eq.~(\ref{BSeigenvalueEquation}),  reserving 
the labels without  primes for the states that result from action
by the Hamiltonian. Thus, the plus and perpendicular components
of fermions momenta are $p'^+=P_{1'2'}^+/2$ and $p'^\perp = 
P_{1'2'}^\perp/2$. In the FF dynamics, we can consider arbitrary 
values of the fermions total momentum components $P_{1'2'}^+$ 
and $P_{1'2'}^\perp$, while the individual fermions' kinematic 
momentum components are always of the form given in 
Eqs.~(\ref{pprime+BS}) and (\ref{pprimeperpBS}), in which 
$x'=x_{1'} = x_{2'} = 1/2$ and $k'^\perp = 0$. The wave 
function $\psi_{t \, 1'2'}$ in Eq.~(\ref{BSeigenvalueEquation}) 
that would correspond to the state $|1'2'\rangle$ would enforce 
with arbitrary accuracy that $(x',k'^\perp)=(1/2,0)$. We illustrate 
the relativistic potentials by results of their action on such wave 
functions.

We extract the relativistic potentials from the matrix elements 
$\langle 1_t 2_t | H_{t \, {\rm eff} \, 2 \, q \bar q} | 1'_t 2'_t\rangle$
in Eq.~(\ref{BSeigenvalueEquation}) one-by-one in the order of lines 
$L_1$ to $L_4$ in Eq.~(\ref{4lines}). We remind the reader that the
coupling constant is factored out. The potentials are functions of 
kinematic components of momenta of the two fermions that enter and 
two fermions that leave the interaction. Together, these are twelve 
arguments. But the total momentum of fermions is conserved and the 
potentials do not depend on it, no matter how large it is. So, they 
are functions of only six variables $x$, $k^\perp$, $x'$ and $k'^\perp$. 
In action on the wave functions $\psi_{t \, 1'2'}$ that we introduced 
above, the primed variables have fixed values $x'=1/2$ and $k'^\perp =0$. 
In addition, as a consequence of rotational symmetry around $z$-axis 
and $k'^\perp$ being zero, the result of action of a potential depends 
only on the variables $x$ and $k^{\perp 2}$. We denote 
\beq
Q \es |k^\perp| \ .
\eeq
This way we obtain four functions $V_1(x,Q)$ to  $V_4(x,Q)$ from
the potentials $V_1(121'2') $ to $V_4(121'2')$ in Eqs.~(\ref{potentialV1}), 
 (\ref{potentialV2}), (\ref{potentialV3}) and (\ref{potentialV4}),
so that for $i = 1$, 2, 3 and 4 we have
\beq
\label{VxQ}
V_i(x,Q) \es V_i(12p'p')  \ .
\eeq
These functions are plotted in comparison with two reference 
functions defined below. The reference functions correspond 
to the intuitive potentials that apply in  non-relativistic quantum 
mechanics.

The first reference function is defined using the momentum 
representation of the attractive Yukawa potential in non-relativistic 
quantum mechanics, which reads
\beq
V_Y(\vec k ,\vec k\,') \es { - g^2 \over (\vec k - \vec k\,' )^2 + \kappa^2 } \ .
\eeq
Since the relative momentum in the state $|1'2'\rangle$
that we use is zero, one sets $\vec k\,'$ to zero. The 
argument of the Yukawa non-relativistic potential 
reduces to $\vec k\,^2$. We identify the non-relativistic 
$\vec k$ with its FF counterpart using formulas of 
App.~\ref{notation}, 
\beq
\vec k\,^2 \es \cM_{12}^2/4 - m^2 \rs { Q^2 + [(x-1/2)2m]^2 \over 4x(1-x) } \ .
\eeq
In the non-relativistic limit the denominator $4x(1-x)$ turns into 1.
Therefore, the Yukawa potential function we could use as a reference
would be
\beq
 {- g^2 \over Q ^2 + [(x-1/2)2m]^2  +  \kappa^2 } \ .
\eeq
However, in lines $L_1$ to $L_4$ we have factored out spinor
matrix elements and the square of the coupling constant with 
proper signs. Our Yukawa reference function is therefore defined 
to be
\beq
\label{VY}
V_Y(x,Q) \es {4m^2 \over Q ^2 + [(x-1/2)2m]^2  +  \kappa^2 } \ .
\eeq
For small $\kappa/m$, the maximal value of this function 
equals $4m^2/\kappa^2$ and the minimal one is zero.

Our second reference function is designed for the annihilation 
channel potentials. We strip the RGPEP form factor from the
potential $V_3$ in Eq.~(\ref{potentialV3}) and obtain
\beq
2m^2 \left(  {1\over b} + {1 \over b'} \right) 
\eeq
with $b = \cM_{12}^2 - \kappa^2$ and $b'=4m^2-\kappa^2$.
Our annihilation reference function is hence defined to be
\beq
\label{VA}
V_A(x,Q) \es 2m^2 \ \left[ { x(1-x) \over Q^2 + m^2  - \kappa^2 x(1-x) } 
+
{ 1 \over 4m^2 - \kappa^2 } \right] \ .
\eeq
Its maximal value is one and it tends to 1/2 for large values 
of $Q$ or extreme values of $x$, when $\kappa \ll m$.

In all figures that illustrate the relativistic potentials, we use 
the same boson mass $\kappa = m/7$ and the same size 
of effective particles $s = (1.5 \, m)^{-1}$. These choices 
are made for purely graphical reasons, to satisfy the condition 
that the characteristic features of the interactions are visible 
well. When the mass $\kappa$ decreases, the Yukawa potential 
at small momentum transfers becomes increasingly spiky and 
approximates the Coulomb potential near $x=1/2$ and $Q=0$
increasingly well. For the parameters chosen in the figures, 
the Yukawa-like  potentials reach the value $4m^2/\kappa^2 = 196$,
see Eq.~(\ref{VY}). When the size $s$ increases, the potentials 
lose strength off shell, which means they are exponentially 
limited to a smaller range of $x$ and $Q$. When $s$ decreases, 
the range increases according to the rule $[Q^2+m^2(2x-1)^2]
/[x(1-x)] \lesssim s^{-2}$.

\begin{figure}[ht!]
\caption{ Relativistic gauge-boson exchange potential 
$V_1(121'2')$ of Eq.~(\ref{potentialV1}). The upper plot
illustrates $V_1(121'2')$ in terms of the potential function 
$V_1(x,Q)$ of Eq.~(\ref{VxQ}). For graphical reasons, the 
boson mass $\kappa$ is set to one seventh of the fermion 
mass $m$ and the RGPEP running size parameter $s$ to 
the inverse of $1.5 \, m$. The variable $x$ corresponds to 
the parton-model $x$ of the fermion labeled by 1. The 
variable $Q$ is the magnitude of transverse momentum of 
that fermion with respect to the anti-fermion labeled by 2. 
The middle panel shows the Yukawa potential function 
$V_Y(x,Q)$ of Eq.~(\ref{VY}), hardly discernible from 
$V_1(x,Q)$. The bottom figure presents the ratio $R_1(x,Q) 
= V_1(x,Q) / V_Y(x,Q)$ of Eq.~(\ref{ratioR1}). The ratio 
exhibits the exponential suppression of effective interactions 
when the invariant mass changes by more than the inverse 
of the RGPEP scale parameter $s$. More details are in the text.}
\label{fig:V1}
\includegraphics[width=.43\textwidth]{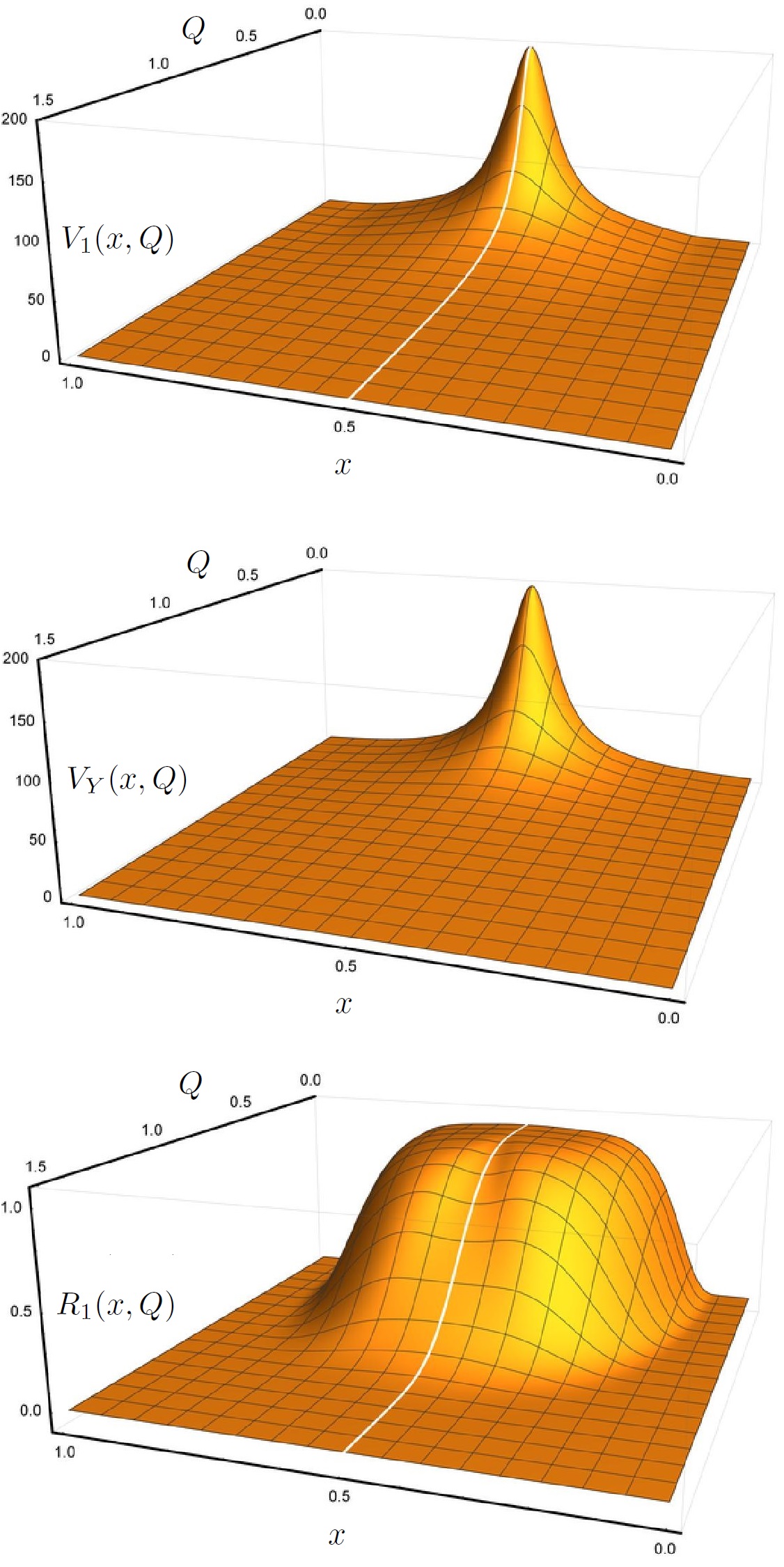}
\end{figure}
Figure~\ref{fig:V1} contains three panels that show, counting from
the top to bottom, the boson exchange potential function $V_1(x,Q)$ 
of Eq.~(\ref{potentialV1}), the Yukawa potential function $V_Y(x,Q)$ 
of Eq.~(\ref{VY}) and their ratio
\beq
\label{ratioR1}
R_1(x,Q)  \es V_1(x,Q) / V_Y(x,Q) \ .
\eeq
These figures demonstrate the role of the RGPEP form factors
in effective interactions. The form factors exponentially suppress
the interactions that change the effective fermions invariant mass 
by more than the inverse of an effective fermion size $s$.
While the relativistic potential function $V_1(x,Q)$ appears almost 
indistinguishable from the Yukawa potential function $V_Y(x,Q)$, 
their ratio displays a huge difference from one, due to the RGPEP 
form factors. In the figures, fermions 1' and 2' have the invariant 
mass squared equal $\cM'^2 = 4m^2$. Fermions 1 and 2 have 
the invariant mass squared equal $\cM^2 = (Q^2 + m^2)/[x(1-x)]$. 
Generally, the RGPEP form factors exponentially suppress the 
interactions off-shell extent according to the rule $(\cM^2 - \cM'^2) 
\lesssim s^{-2}$.  When the variable $x'$ introduced below 
Eq.~(\ref{state1'2'}) deviates from 0.5, the Yukawa peak of 
Fig.~\ref{fig:V1} shifts and centers on $x = x'$ instead of 0.5. 
If the transverse momentum $k'^\perp$ significantly differs from 
zero, the potential function behaves in a somewhat more complicated 
way due to its additional dependence on $x'$, $Q'$ and the angle 
between $k^\perp$ and $k'^\perp$ in the transverse plane, but it 
follows the rule that $(Q^2+m^2)/[x(1-x)]$ does not differ from 
$(Q'^2+m^2)/[x'(1-x')]$ by much more than $1/s^2$.

\begin{figure}[h!]
\caption{Relativistic potential $V_2(121'2')$ of Eq.~(\ref{potentialV2}).
It is drawn in terms of the potential function $V_2(x,Q)$ of  Eq.~(\ref{VxQ}), 
in orange. For comparison, the Yukawa-like potential function $V_1(x,Q)$ 
of Eq.~(\ref{VxQ}), see Fig.~\ref{fig:V1}, is shown in blue. The functions 
are displayed with the same sign to show their relative magnitudes well. 
The view of potentials is arranged to be from the opposite point to that 
in Fig.~\ref{fig:V1} in order to show the relative magnitude of the two 
functions at small momentum transfers, which is the region where the 
bound-state formation mechanism is most active. In that region, the 
potential function $V_2(x,Q)$ is much smaller in size than the Yukawa-like
function $V_1(x,Q)$. The Yukawa peak reaches $4 \times 49$, as explained 
below Eq.~(\ref{VY}). The potential function $V_2(x,Q)$ vanishes at that point.}
\label{fig:V2}
\includegraphics[width=.43\textwidth]{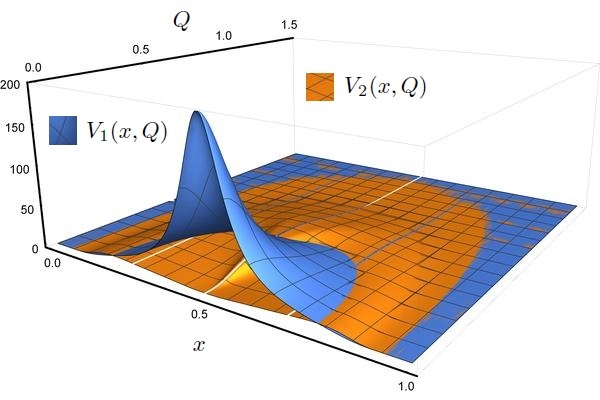}
\end{figure}

Figure~\ref{fig:V2} shows the relativistic FF potential of Eq.~(\ref{potentialV2}) 
in terms of the function $V_2(x,Q)$ in Eq.~(\ref{VxQ}), in comparison with the 
potential function $V_1(x,Q)$, shown in Fig.~\ref{fig:V1} for the same parameters 
$\kappa$ and $s$. It is visible that the relativistic FF potential $V_2(121'2')$ has
support only off shell. In the region of binding, it is very small in comparison
to the one-boson-exchange potential $V_1(121'2')$. In the language of 
SRG~\cite{GlazekWilson}, it has significant matrix elements only outside the 
band of a band-diagonal matrix of the effective Hamiltonian, whose width 
in terms of the invariant mass is $1/s$. Far away from the diagonal, the function 
$V_2(x,Q)$ briefly exceeds the function $V_1(x,Q)$, where the latter is already 
two orders of magnitude smaller than in the band. This partial dominance of 
$V_2(x,Q)$ over $V_1(x,Q)$ is the origin of the huff-like pattern visible in 
Fig.~\ref{fig:V2}.

The potential $V_2$ does not contribute to the on-shell scattering matrix 
in the Born approximation and does not have a classical counterpart, in 
contrary to the potential $V_1$ that corresponds to the Yukawa potential. 
This is a welcome feature because the potential $V_2$ multiplies the 
non-covariant spin structure $EX_+$ of Eq.~(\ref{EX+}), see Eq.~(\ref{L2EX+V2}). 
The factor $EX_+$ preserves spins of fermions and introduces the factor 
$\sqrt{x(1-x)x'(1-x')}$ that further suppresses the interaction for extreme values 
of $x$ or $x'$. The alien feature of $V_2(x,Q)$ near $x=1/2$ originates from 
the factor $1/(x-x')$ that produces a discontinuous variation of the potential 
as a function of $x$ for $x'\neq 1/2$. The discontinuity is suppressed by 
additional powers of $x-1/2$ for $x'=1/2$. For $x'\neq 1/2$, it is integrable 
with regular wave functions of $x$ and $x'$ in the sense of principal value.

\begin{figure}[h!]
\caption{Relativistic annihilation-channel potential $V_3(121'2')$ 
of Eq.~(\ref{potentialV3}), shown in terms of the potential function 
$V_3(x,Q)$ of Eq.~(\ref{VxQ}). One sees the effect of the RGPEP
form factors. The middle panel shows the annihilation-channel
potential function $V_A(x,Q)$ of Eq.~(\ref{VA}). The ratio 
$R_3(x,Q)=V_3(x,Q)/V_A(x,Q)$ is shown in the bottom panel, see the text.} 
\label{fig:V3}
\includegraphics[width=.43\textwidth]{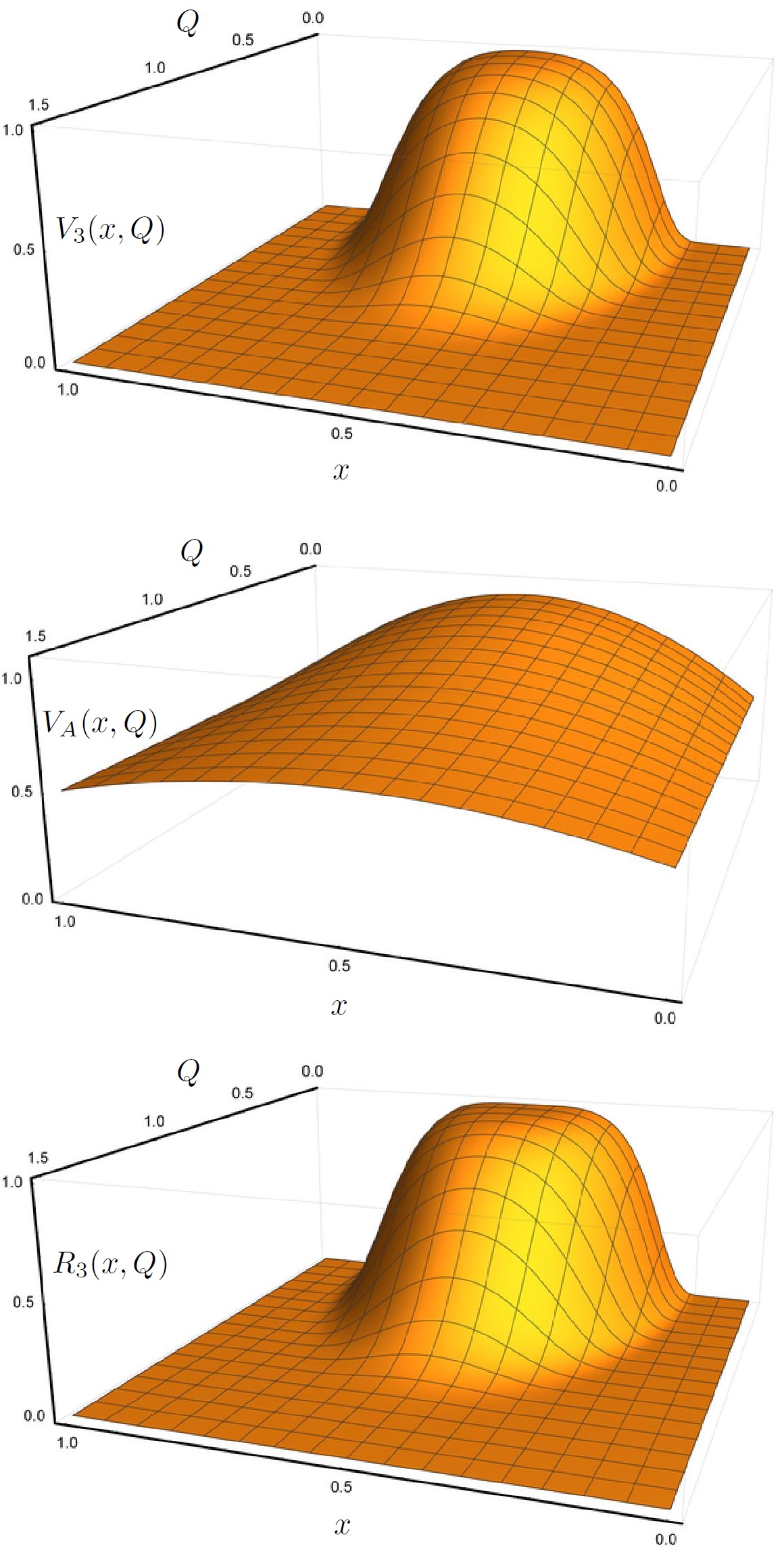}
\end{figure}
Figure~\ref{fig:V3} shows three panels that, counting from 
the top to botom, illustrate the boson annihilation channel 
potential function $V_3(121'2')$ of Eq.~(\ref{potentialV3}).
The top panel shows the function $V_3(x,Q)$ of Eq.~(\ref{VxQ}).
The middle panel shows the potential function $V_A(x,Q)$ of 
Eq.~(\ref{VA}). The ratio $R_3(x,Q)  = V_3(x,Q) / V_A(x,Q)$
is shown in the bottom panel. Comparing the panels top with 
middle, one sees again the role of the RGPEP form factors. 
In the SRG language, they squeeze the potential to the band 
of effective theory. The bottom-panel ratio function $R_3(x,Q)$ 
is characterized by a little more flat shape than the top
panel potential function $V_3(x,Q)$. This effect shows that
the RGPEP form factor introduces a relativistic annihilation-channel 
potential that is close to the function $V_A(x,Q)$ times the RGPEP
form factor.

\begin{figure}[h!]
\caption{ Relativistic annihilation-channel potential functions $-10 \, V_4(x,Q)$ and $V_3(x,Q)$.
See the text for details.}
\label{fig:V4}
\includegraphics[width=.43\textwidth]{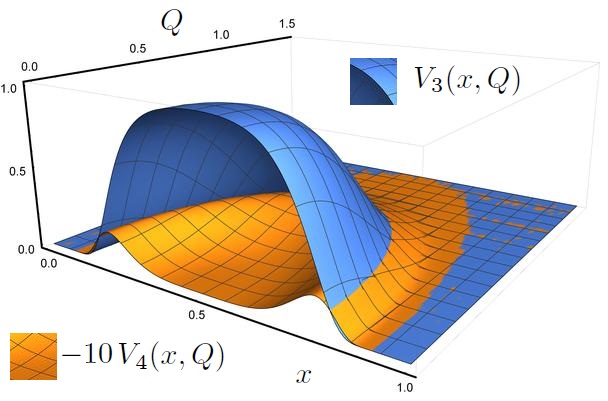}
\end{figure}
Finally, Fig.~\ref{fig:V4} illustrates the relativistic FF annihilation-channel 
potential $V_4(121'2')$ of Eq.~(\ref{potentialV4}) in terms of the 
potential function $V_4(x,Q)$ in Eq.~(\ref{VxQ}), shown simultaneously 
with the potential function $V_3(x,Q)$ of Eq.~(\ref{VxQ}). The comparison
shows the smallness of $V_4(x,Q)$. Its sign is changed and its value is 
multiplied by ten in order to obtain an informative picture. The relativistic 
potential $V_4(121'2')$ appears in the line $L_4$ in Eq.~(\ref{L4AN+V4}) 
multiplied by the frame-dependent spin factor $AN_+$ of Eq.~(\ref{AN+}). 
It does not contribute to the on-shell scattering matrix in the Born 
approximation and does not have any familiar counterpart in quantum 
mechanics. However, it does participate in the off-shell bound-state dynamics, 
in addition to the potential $V_3$. Its significance in that dynamics is 
not known at this point. The actual magnitude of the boson mass $\kappa$ 
much smaller than the fermion mass $m$, does not influence the potentials 
$V_3$ and $V_4$ in any significant way. 

\section{ Conclusion }
\label{conclusion}

The RGPEP allows one to calculate second-order effective 
masses and interactions in the fermion-anti-fermion systems 
in Abelian gauge theory. 
Canonical Hamiltonian leads to 
difficulties with unambiguous handling of small $x$ and 
large $k^\perp$ singularities because the singular terms 
involve the ratio $k^{\perp 2}/x$ and the ultraviolet 
divergences are mixed with the small $x$ divergences.
As a result, the ultraviolet 
counter terms involve unknown functions of $x$ and small-$x$ 
counter terms contain functions of $k^\perp$~\cite{Wilsonetal}. 
However, once the mass parameter for gauge bosons is 
introduced according to the principles of local gauge symmetry 
and spontaneous violation of the global gauge symmetry, 
a mass gap is introduced and one achieves unambiguous 
control on the divergences. The ultraviolet, small-$x$ and 
infrared singularities are separated from each other in a way 
specific to the FF Hamiltonian dynamics and the RGPEP 
evolution of Hamiltonian operators. Namely, the longitudinal 
small-$x$ region is controlled by the parameter $s\kappa$ 
while the transverse ultraviolet region is controlled by $s$, 
where $s$ is the RGPEP {\it a priori} arbitrary scale parameter. 
The origin of the separation lies in the expression 
\beq
s^2 \delta \cM^2  \es  {(s\kappa)^2 + (sk^\perp)^2 \over x}
\eeq
for the contribution of bosons to the arguments of exponentially 
falling-off RGPEP form factors in the effective interactions. 
It is visible that one cannot make $s^2\cM^2$ small for small 
$x$ by making $sk^\perp$ small because eventually $s\kappa$ 
begins to count and $s\cM$ always diverges for fixed $s$ when 
$x \to 0$.

Using expansion in the coupling constant $g$, one can employ
the RGPEP to study what happens when the boson mass 
$\kappa$ is varied and what comes out in terms of the 
effective theory when $\kappa$ is made very small. The 
result of second-order calculations described in this article 
is that the fermion mass counter terms can reach enormous 
values. Their contribution is canceled precisely in the second-order
mass eigenvalue equation for physical fermions or bosons, 
but the canceled terms are much greater than the eigenvalues, 
if the size of effective quanta is very small. However, when 
that size is increased toward the fermions Compton wavelength 
and above, the mass corrections become very small. 

It is also found that the effective boson masses vary differently 
with the size $s$ for the commonly known transverse bosons 
and for the less known longitudinal ones. The mass corrections 
for the latter stay small or very small in comparison to the mass 
corrections for the former.

The RGPEP also allows one to calculate interaction terms that 
drive the fermion-anti-fermion bound state dynamics. One 
obtains Yukawa potentials that tend to the Coulomb potential 
when the boson mass tends to zero and the size of effective 
quanta increases to and above the fermion Compton wavelength. 
However, the size increase is associated with development of 
increasingly important vertex form factors that suppress 
interactions with large changes of the invariant mass of 
fermions.

The fact that the FF Hamiltonian dynamics is invariant with
respect to the Lorentz boosts along one axis, besides six
other Poincar\'e transformations, allows one to relate the
RGPEP results for the Coulomb- or Yukawa-like systems 
to their parton model picture. The results described in this 
article suggest that when we imagine partons as constituents, 
their size cannot be ignored. If one ignores their size, the 
power-like behavior of perturbative interactions is extended 
to the phase-space region where the eigenvalue condition 
for bound states imposes decisive departures of the wave 
functions from their perturbative estimates that use canonical
interactions. The effective interactions become exponentially 
suppressed when the fermions invariant mass changes by 
more than the inverse of their Compton wavelength. 
One also obtains small effective interaction terms that 
appear in addition to the Coulomb and Yukawa potentials  
and do not have classical counterparts. The RGPEP enables
us to draw details of all these potentials. 

It is not clear what happens in the higher order RGPEP 
calculations. Of key interest is the fourth order. This is where
the running of effective coupling constant shows up in the
bound-state dynamics for the first time. The computation 
is certainly doable and the results would be of interest.

The final question we wish to address is whether Soper's 
theory is a valid approximation to the gauge theory with 
spontaneously broken global symmetry. We obtain the 
former from the latter in the massive limit in which the 
classical field $h/v$ is set to zero when $v$ is formally 
sent to infinity. However, the limit is considered in a classical 
Lagrangian. The effective quantum theory derived using 
the RGPEP will include corrections that depend on the 
momentum range $1/s_r$ of interactions in ratio to $v$. 
The order of limits $v \to \infty$ and $s_r \to 0$ may matter.  
At this point, the calculations described here are considered 
reasonable regarding gauge symmetry because they are 
carried out using the massive limit that results in the Soper 
theory, which by itself is an example of a theory with a 
form of gauge symmetry. The full theory, not using the 
massive limit, can also be analyzed using the RGPEP. 

\begin{appendix}

\section{ Notation}
\label{notation}

Translation invariance on the front implies conservation
of momentum described by the $\delta$-function
$\tilde \delta_{c.a}$, where $c$ denotes created and 
$a$ annihilated quanta. We use the convention
\beq
\tilde \delta_{c.a}
\es
2 (2\pi)^3 \delta(P_c^+ - P_a^+) \delta^{2}(P_c^\perp - P_a^\perp) \ ,
\eeq
where $P_c$ and $P_a$ denote the total momenta of particles
created and annihilated, respectively. The corresponding invariant 
masses are $\cM_c^2 = P_c^2$ and $\cM_a^2 = P_a^2$
with minus components of individual particles momenta
calculated from their mass shell conditions, $p^- = 
(m^2 + p^{\perp 2})/p^+$.

Integration over a single particle phase space,  
\beq
\int d^4p \ \delta(p^2-m^2) \ \theta(p^0)
\es
\int { d^3 p \over 2 E_p} 
\rs 
\int_0^\infty { dp^+ \over 2 p^+} \int d^2 p^\perp \ ,  
\eeq
is denoted by $\int[p]$ and if one has more
particles to integrate over their momenta
$p_1$, $p_2$, . . . $p_n$, the integral is 
abbreviated to
\beq
\int[12 . . . n] \es \int[p_1] \int[p_2]. . .\int[p_n] \ .
\eeq
When two particles have together momentum $P$
and carry fractions $x$ and $1-x$ of it and some 
transverse relative momentum $k^\perp$,
\beq
p_1^+ \es x P^+ \ , \\
p_2^+ \es (1-x) P^+ \ , \\
p_1^\perp \es x P^\perp + k^\perp \ , \\
p_2^\perp \es (1-x) P^\perp - k^\perp \ ,
\eeq
one has
\beq
\label{momentumP12}
\int [12] \es \int [P] \int [xk]  \ , \\
\label{momentumk12}
\int[xk] \es \int_0^1 { dx \over 4\pi x(1-x)}  \int {d^2 k^\perp \over (2\pi)^2 } \ .
\eeq
In terms of the relative three-momentum of two 
particles of mass $m$ in their rest frame, $\vec k$, 
\beq
x     \es (1 + k_z/E_k)/2 \ , \\
{dx \over x(1-x) } \es { 2 \ dk_z \over E_k} \ , \\
\label{minimalrelativity}                             
\int[xk] \es \int { d^3 k \over (2\pi)^3  E_k}  \ ,
\eeq
where $E_k = \sqrt{ m^2 + \vec k\,^2}$.
The invariant mass of two particles is
\beq
(p_1+p_2)^2 \es (p_1^+ + p_2^+) (p_1^- + p_2^-) - (p_1^\perp + p_2^\perp)^2 \\
                    \es {k^{\perp 2} + m_1^2 \over x} + {k^{\perp 2} + m_2^2 \over 1-x}  \\
                    \es \left( \sqrt{ m_1^2 + \vec k\,^2} + \sqrt{ m_2^2 + \vec k\,^2} \right)^2 \ .
\eeq
We use spinors $ u_{p\sigma} = B(p,m) u_\sigma$ and 
$v_{p\sigma} = B(p,m) v_\sigma$ in which the spinors 
at rest are related by $v_\sigma = C u^*_\sigma$
with $C = i\gamma^2$ and the front boost matrix is 
$B(p,m) = {1\over \sqrt{p^+ m}} [ \Lambda_+ p^+ + \Lambda_- (m + p^\perp \alpha^\perp)]$,
where $\Lambda_\pm = {1\over 2}(1 \pm \alpha^3) $. 
The spinors at rest are
\beq 
u_\sigma \es \sqrt{2m}\left[ \begin{array}{c} \chi_\sigma \\ 0
\end{array}\right] \ , \quad
  v_\sigma \rs 
\sqrt{2m_f}\left[ \begin{array}{c} 0 \\ \xi_{-\sigma}
\end{array}\right] \ ,
\eeq
where
$  \xi_{-\sigma}=-i\sigma_2\chi_\sigma = \sigma\chi_{-\sigma}$,
{\it cf.}~\cite{LepageBrodsky,fermions}. Free bosons of type $A$ 
have polarization vectors 
\beq
\varepsilon^\mu_{p \sigma} 
\es 
\left( \varepsilon^-_{p\sigma} 
= 2p^\perp \varepsilon^\perp_\sigma/p^+,  
\varepsilon^\perp_\sigma \right) 
\eeq
with $\varepsilon^\perp_\sigma = (1 + \sigma, 1-\sigma)/2$.
Free bosons of type $B$ have polarization vectors 
\beq
\varepsilon_{p3} \es \left( \varepsilon_{p3}^- = {p^{\perp \, 2}
- \kappa^2 \over \kappa p^+}, ~
\varepsilon_{p3}^+ = {p^+      \over \kappa}, ~\varepsilon_{p3}^\perp = {p^\perp \over \kappa} \right) \\\es  {p \over \kappa}  - \eta {\kappa \over p^+} \ ,\eeq
and $\eta^+=\eta^\perp = 0$ while $\eta^- = 2$.

\section{ Details of the initial Hamiltonian }
\label{hamiltonianappendix}

The canonical Hamiltonian terms in Eq.~(\ref{initialcH})
are listed below using notation explained in App.~\ref{notation}. 
The subscript 0 associated with canonical creation and 
annihilation operators for the bare quanta that are 
considered point-like, or of size $s_r = t_r^{1/4} \to 
0$ as the regularization is being lifted, is not needed 
here and it is omitted. The free part of the Hamiltonian 
is $H_f = H_{\psi^2} + H_{A^2} + H_{B^2}$, where
\beq
\label{Hpsi2app}
H_{\psi^2} 
\es 
\sum_{\sigma = 1}^2 \int [p] {p^{\perp \, 2} + m^2 \over p^+}
  \left[b^\dagger_{p\sigma }b_{p\sigma } + 
  d^\dagger_{p\sigma }d_{p\sigma } \right] \ , \\
\label{HA2app}
H_{A^2} 
\es 
\sum_{\sigma =1}^2 \int [p] {p^{\perp \, 2} + \kappa^2 \over
p^+}
  a^\dagger_{p\sigma}a_{p\sigma} \ , \\
\label{HB2app}
H_{B^2} 
\es 
\int [p] {p^{\perp \, 2} + \kappa^2 \over p^+} 
  c^\dagger_p c_p \ .
\eeq
The interaction Hamiltonian $H_I = H - H_f$ contains 
terms of orders $g$ and $g^2$. The terms order $g$ 
are
\beq
\label{HpsiApsi}
H_{\psi A \psi} 
\es   
g \sum_{123}\int[123] \,
  \tilde \delta_{c.a} \, 
  \left[
\bar u_2 \hspace{-3pt} \not\!\varepsilon_1^* u_3 \ b^\dagger_2
a^\dagger_1 b_3
- \bar v_3 \hspace{-3pt} \not\!\varepsilon_1^* v_2 \ d^\dagger_2
a^\dagger_1 d_3
+\bar u_1 \hspace{-3pt} \not\!\varepsilon_3 v_2 \ b^\dagger_1
d^\dagger_2 a_3 \ + \ h.c.
  \right] 
\ , \\
\label{HpsiBpsi}
H_{\psi B \psi}
\es
- g \sum_{12}\int[123] \,
  \tilde \delta_{c.a} \,  
  \left[
\bar u_2 { \kappa \gamma^+ \over p_1^+} u_3 \ b^\dagger_2
c^\dagger_1 b_3
- \bar v_3 { \kappa \gamma^+ \over p_1^+} v_2 \ d^\dagger_2
c^\dagger_1 d_3
+\bar u_1 { \kappa \gamma^+ \over p_3^+} v_2 \ b^\dagger_1
d^\dagger_2 c_3 \ + \ h.c.
  \right]          \ .
\eeq
There are two terms order $g^2$. The term 
due to constraint on $\psi_-$ is
\beq
H_{\psi A A \psi} 
\es   
{g^2 \over 2}
\sum_{1234}\int[1234] \,
\tilde \delta_{c.a} \, \left\{ ~
\right\}_{\psi A A \psi} \ ,
\eeq
where, in the universal order $b^\dagger d^\dagger a^\dagger a d b$,
\beq
\label{pAAp}
\left\{ ~ \right\}_{\psi A A \psi} 
\es
{  \bar u_1 \hspace{-3pt} \not\!\varepsilon_2^* 
                \gamma^+ 
                \hspace{-5pt} \not\!\varepsilon_3 u_4  
                \over p_3^+ + p_4^+ } 
                \ b_1^\dagger a_2^\dagger a_3 b_4   
+
{  \bar u_1 \hspace{-3pt} \not\!\varepsilon_2^* 
                \gamma^+ 
                \hspace{-5pt} \not\!\varepsilon_3 v_4  
                \over p_3^+ - p_4^+ } 
                \ b_1^\dagger d_4^\dagger a_2^\dagger a_3  
+
{  \bar u_1 \hspace{-3pt} \not\!\varepsilon_2^* 
                \gamma^+ 
                \hspace{-5pt} \not\!\varepsilon_3^* u_4  
                \over  p_4^+ - p_3^+ } 
                \ b_1^\dagger a_2^\dagger a_3^\dagger b_4 
\np
{  \bar u_1 \hspace{-3pt} \not\!\varepsilon_2
                \gamma^+ 
                \hspace{-5pt} \not\!\varepsilon_3 u_4  
                \over p_3^+ + p_4^+ } 
                \ b_1^\dagger a_2 a_3 b_4   
+
{  \bar u_1 \hspace{-3pt} \not\!\varepsilon_2
                \gamma^+ 
                \hspace{-5pt} \not\!\varepsilon_3 v_4  
                \over p_3^+ - p_4^+ } 
                \ b_1^\dagger d_4^\dagger a_2 a_3 
+
{  \bar u_1 \hspace{-3pt} \not\!\varepsilon_2
                \gamma^+ 
                \hspace{-5pt} \not\!\varepsilon_3^* u_4  
                \over  p_4^+ - p_3^+ } 
                \ b_1^\dagger a_3^\dagger a_2 b_4 
-
{  \bar u_1 \hspace{-3pt} \not\!\varepsilon_2
                \gamma^+ 
                \hspace{-5pt} \not\!\varepsilon_3^* v_4  
                \over p_3^+ + p_4^+ } 
                \ b_1^\dagger d_4^\dagger a_3^\dagger a_2  
\np
{  \bar v_1 \hspace{-3pt} \not\!\varepsilon_2^* 
                \gamma^+ 
                \hspace{-5pt} \not\!\varepsilon_3 u_4  
                \over p_3^+ + p_4^+ } 
                \  a_2^\dagger a_3 d_1 b_4   
+
{  \bar v_1 \hspace{-3pt} \not\!\varepsilon_2^* 
                \gamma^+ 
                \hspace{-5pt} \not\!\varepsilon_3 v_4  
                \over  p_4^+ - p_3^+  } 
                \ d_4^\dagger a_2^\dagger a_3 d_1
+
{  \bar v_1 \hspace{-3pt} \not\!\varepsilon_2^* 
                \gamma^+ 
                \hspace{-5pt} \not\!\varepsilon_3^* u_4  
                \over  p_4^+ - p_3^+ } 
                \  a_2^\dagger a_3^\dagger d_1 b_4 
+
{  \bar v_1 \hspace{-3pt} \not\!\varepsilon_2^* 
                \gamma^+ 
                \hspace{-5pt} \not\!\varepsilon_3^* v_4  
                \over p_3^+ + p_4^+ } 
                \ d_4^\dagger  a_2^\dagger a_3^\dagger d_1
\np
{  \bar v_1 \hspace{-3pt} \not\!\varepsilon_2
                \gamma^+ 
                \hspace{-5pt} \not\!\varepsilon_3 v_4  
                \over  p_4^+ - p_3^+} 
                \ d_4^\dagger  a_2 a_3 d_1
+
{  \bar v_1 \hspace{-3pt} \not\!\varepsilon_2
                \gamma^+ 
                \hspace{-5pt} \not\!\varepsilon_3^* u_4  
                \over p_4^+ - p_3^+  } 
                \   a_3^\dagger a_2 d_1 b_4 
+
{  \bar v_1 \hspace{-3pt} \not\!\varepsilon_2
                \gamma^+ 
                \hspace{-5pt} \not\!\varepsilon_3^* v_4  
                \over p_3^+ + p_4^+ } 
                \ d_4^\dagger  a_3^\dagger a_2  d_1 \ .
\eeq
The term due to constraint on $A^-$ is 
\beq
H_{(\psi\psi)^2}
\es
{g^2 \over 2}
\sum_{1234}\int[1234] \,
\tilde \delta_{c.a} \,  \left\{ ~
\right\}_{(\psi\psi)^2} \ ,
\eeq
where $\left\{ ~ \right\}_{(\psi\psi)^2}$ reads
\beq
\label{Hpsi2ordered}
\left\{ ~ \right\}_{(\psi\psi)^2}
\es
-
{ \bar u_1 \gamma^+ u_2 \ \bar u_3 \gamma^+ u_4 \over (p_3^+ -
p_4^+)^2 }
\
b_1^\dagger  b_3^\dagger b_2 b_4  
+
{ \bar u_1 \gamma^+ u_2 \ \bar u_3 \gamma^+ v_4 \over (p_3^+ +
p_4^+)^2 }
\
b_1^\dagger  b_3^\dagger d_4^\dagger  b_2
\nm
{ \bar u_1 \gamma^+ u_2 \ \bar v_3 \gamma^+ u_4 \over (p_3^+ +
p_4^+)^2 }
\
b_1^\dagger  d_3 b_2 b_4  
-
{ \bar u_1 \gamma^+ u_2 \ \bar v_3 \gamma^+ v_4 \over (p_3^+ -
p_4^+)^2 }
\
b_1^\dagger  d_4^\dagger   d_3 b_2 
\nm
{ \bar u_1 \gamma^+ v_2 \ \bar u_3 \gamma^+ u_4 \over (p_3^+ -
p_4^+)^2 }
\
b_1^\dagger b_3^\dagger  d_2^\dagger b_4  
\np
{ \bar u_1 \gamma^+ v_2 \ \bar v_3 \gamma^+ u_4 \over (p_3^+ +
p_4^+)^2 }
\
b_1^\dagger d_2^\dagger d_3 b_4  
-
{ \bar u_1 \gamma^+ v_2 \ \bar v_3 \gamma^+ v_4 \over (p_3^+ -
p_4^+)^2 }
\
b_1^\dagger d_2^\dagger  d_4^\dagger  d_3
\np
{ \bar v_1 \gamma^+ u_2 \ \bar u_3 \gamma^+ u_4 \over (p_3^+ -
p_4^+)^2 }
\
b_3^\dagger d_1 b_2  b_4  
+
{ \bar v_1 \gamma^+ u_2 \ \bar u_3 \gamma^+ v_4 \over (p_3^+ +
p_4^+)^2 }
\
 b_3^\dagger d_4^\dagger  d_1 b_2
\np
{ \bar v_1 \gamma^+ u_2 \ \bar v_3 \gamma^+ v_4 \over (p_3^+ -
p_4^+)^2 }
\
d_4^\dagger  d_1  d_3  b_2
\nm
{ \bar v_1 \gamma^+ v_2 \ \bar u_3 \gamma^+ u_4 \over (p_3^+ -
p_4^+)^2 }
\
b_3^\dagger d_2^\dagger  d_1 b_4  
+
{ \bar v_1 \gamma^+ v_2 \ \bar u_3 \gamma^+ v_4 \over (p_3^+ +
p_4^+)^2 }
\
 b_3^\dagger d_2^\dagger  d_4^\dagger  d_1
\nm
{ \bar v_1 \gamma^+ v_2 \ \bar v_3 \gamma^+ u_4 \over (p_3^+ +
p_4^+)^2 }
\
 d_2^\dagger d_1 d_3 b_4  
-
{ \bar v_1 \gamma^+ v_2 \ \bar v_3 \gamma^+ v_4 \over (p_3^+ -
p_4^+)^2 }
\
 d_2^\dagger d_4^\dagger  d_1 d_3 
\ .
\eeq

\subsection{ Regularization }
\label{Appregularization}

Both Hamiltonian terms $H_{\psi A A \psi} $ and $H_{(\psi
\psi)^2}$
contain a product of four bare Fock operators corresponding 
to two factors $h_{12}$ and $h_{34}$ and inverse of 
$i \partial^+$ or $(i \partial^+)^2$,
\beq
h_{12}  {1 \over (i \partial^+)^n } h_{34}
\eeq
with $n=1$ or $n=2$. In agreement with their 
origin in constraints, the operators $h_{12}$ 
and $h_{34}$ are regulated as the operators 
order $g$ are through the RGPEP vertex form 
factors with the size parameter $s_r = t_r^{1/4}$,
see Eqs.~(\ref{HpsiApsi2}) and (\ref{HpsiBpsi2})
and comments below them.

\end{appendix}


\end{document}